\documentclass[a4paper,11pt]{article}
\pdfoutput=1 % if your are submitting a pdflatex (i.e. if you have
\usepackage{jheppub}
\usepackage[T1]{fontenc} % if needed
\usepackage[all]{xy}
\usepackage{rotating}
\usepackage{float}
\usepackage{tikz}
\usepackage{circuitikz}
\usepackage{tikz-network}
\usepackage{diagbox}
\usepackage{braket}
\usepackage{empheq}
\usepackage{tabularx}
\usepackage{multirow}
\usepackage{slashed}
\usetikzlibrary{decorations.pathreplacing}

\usetikzlibrary{decorations.markings}
\usetikzlibrary{arrows.meta}
\tikzset{->-/.style={decoration={
  markings,
  mark=at position #1 with {\arrow{>}}},postaction={decorate}}}

\pgfarrowsdeclare{:}{:}{}{}

\let\a=\alpha \let\b=\beta \let\g=\gamma \let\d=\delta 
  \let\th=\theta  \let\k=\kappa
\let\l=\lambda     \let\r=\rho
\let\s=\sigma \let\t=\tau    
    \let\D=\Delta

\DeclareMathOperator{\Tr}{Tr}

\def\a{\alpha}
\def\b{\beta}

\def\CA{{\cal A}}
\def\CB{{\cal B}}
\def\CC{{\cal C}}

\def\CN{{\cal N}}

\def\CZ{{\cal Z}}

\def\Wm{\widetilde{m}}

\def\Wq{\widetilde{q}}

\def\WN{\widetilde{N}}

\def\WQ{\widetilde{Q}}

\def\beq#1\eeq{\begin{align}#1\end{align}}

\newcommand{\nn}{\nonumber}

\makeatletter
\newcommand*{\rom}[1]{\expandafter\romannumeral #1}

\makeatother%For Roman Numeral

\title{Monopole-deformed Kim--Park dualities \\
from elementary dualities}

\abstract{
We derive the ellipsoid partition function identities for the monopole-deformed Kim–Park dualities of 3d $\mathcal{N}=2$ adjoint SQCD with unitary gauge groups. The derivation uses only Aharony and Benini–Benvenuti–Pasquetti dualities as building blocks. By deconfining the adjoint into a $C_p[U(N)]$ quiver tail, we construct explicit duality sequences. Confinement and a rank-reducing duality chain then yield the expected magnetic partition functions, including the singlet contributions. We also present an alternative partition-function derivation of the $C_p[U(N)]$ s-confinement and provide {\tt Quiver Lab}, a standalone HTML application for implementing and visualizing duality moves on quiver diagrams.
}

\author{Sungjoon Kim,}
\author{Jehyun Lee}

\affiliation[]{Department of Physics, Korea Advanced Institute of Science and Technology, Daejeon 34141, Korea}

\emailAdd{sungjoon@kaist.ac.kr}
\emailAdd{2wpgus@kaist.ac.kr}

\begin{document} 
\maketitle
\flushbottom

%%%%%%%%%%%%%%%%%%%%%%%%%%%%%%%%%%%%%%%%%%%%%%%%%%%%%%%%%%%%%%%%%%%%%%%%
%%%%%%%%%%%%%%%%%%%%%%%%%%%%%%%%%%%%%%%%%%%%%%%%%%%%%%%%%%%%%%%%%%%%%%%%

%%%%%%%%%%%%%%%%%%%%%%%%%%%%%%%%%%%%%%%%%%%%%%%%%%%%%%%%%%%%%%%%%%%%%%%%

\section{Introduction}
Duality has been one of the most intriguing, yet mysterious structures in quantum field theory~(QFT). In particular, numerous examples have been discovered in supersymmetric settings and confirmed using supersymmetric localization techniques~\cite{Seiberg:1994pq,Kutasov:1995np,Brodie:1996vx,Giveon:2008zn,Benini:2011mf,Kapustin:2011gh,Hwang:2011qt,Hwang:2011ht,Hanany:2015via,Hwang:2018uyj,Hwang:2020ddr,Benvenuti:2024mpn,Benvenuti:2024seb,Benvenuti:2025huk,Benvenuti:2025qnq,Hayashi:2025guk,Okazaki:2021pnc,Okazaki:2021gkk,Cho:2026tqf,Benvenuti:2026usm}. 4d $\CN=1$ Seiberg duality is the most pioneering instance as a generalization of electromagnetic duality in SQCD~\cite{Seiberg:1994pq}. More precisely, two distinct UV descriptions of SQCD flow to the same IR fixed point, and are therefore said to be IR dual.

Three dimensions are particularly interesting since gauge theories admit monopole operators, which can be used to deform the theory and generate new dualities~\cite{Benvenuti:2016wet,Amariti:2018gdc,Amariti:2018wht,Amariti:2019rhc,Hwang:2022jjs,Jia:2025koz}. Aharony duality~\cite{Aharony:1997gp} is the 3d $\CN=2$ counterpart of Seiberg duality, and a monopole superpotential deformation gives rise to its cousin, the Benini--Benvenuti--Pasquetti (BBP) duality~\cite{Benini:2017dud}. The Kim--Park duality~\cite{Kim:2013cma} --- the 3d counterpart of the 4d $\CN=1$ Kutasov--Schwimmer duality~\cite{Kutasov:1995np} --- extends Aharony duality by including adjoint matter. The presence of adjoint matter allows for various dressed monopole operators, which in turn give rise to many new dualities upon monopole superpotential deformations~\cite{Hwang:2022jjs}. Motivated by this, a new duality for 3d $\CN=2$ adjoint SQCD with a superpotential containing a single dressed monopole operator $\hat V_\b^\pm$ was proposed in~\cite{Jia:2025koz}, which we refer to as the $KP_\b^\pm$ duality throughout this paper. Together with the original Kim–Park duality, this family accounts for the monopole-deformed dualities classified in~\cite{Jia:2025koz}, up to quadratic powers of monopole operators.

On the other hand, the s-confinement~\cite{Csaki:1996sm} of the 4d $\CN=2$ $D_p[SU(N)]$ Argyres--Douglas theory into adjoint matter was observed in~\cite{Maruyoshi:2023mnv}; see also~\cite{Benvenuti:2018bav,Amariti:2018wht,Nii:2019ebv,Benvenuti:2021nwt,Amariti:2022wae,Bajeot:2023gyl,Amariti:2023wts,Amariti:2024gco,Amariti:2025lem}. Its 3d counterpart was subsequently used to derive the Kim--Park duality from Aharony duality alone~\cite{Hwang:2024hhy,Benvenuti:2024glr}. Moreover, the 3d confinement itself decomposes into a sequence of Aharony and BBP dualities, serving as elementary building blocks in the spirit of dualities from dualities~\cite{Pasquetti:2019uop,Pasquetti:2019tix,Benvenuti:2020gvy,Benvenuti:2020wpc,Bottini:2022vpy,Amariti:2022iaz}; see also~\cite{Bottini:2021vms,Hwang:2021ulb,Comi:2022aqo,Comi:2026lcj} for related developments in algorithmic mirror symmetry.

In this paper, we demonstrate the $KP_\b^\pm$ duality by matching ellipsoid partition functions using s-confinement and sequential dualizations. We deconfine the adjoint matter on one side, then apply a sequence of Aharony and BBP dualities to derive the partition function agreement. We present an alternative derivation of the $\CC_p[U(N)]$ s-confinement of~\cite{Benvenuti:2024glr} and a generalized rank-reducing duality chain in Appendices~\ref{app: conf derivation} and~\ref{app: rr}, respectively. Finally, we provide {\tt Quiver Lab}, a tool for implementing and visualizing duality moves on quiver diagrams; see Appendix~\ref{app: quiver lab}. All these results follow solely from Aharony and BBP dualities, further reinforcing the spirit of dualities from dualities. This raises the intriguing question of whether more general 3d dualities can be derived solely from these two building blocks, which we leave for future work.

\subsection{3d \texorpdfstring{$\CN=2$}{N=2} IR dualities}
We begin with a minimal review of the two 3d $\CN=2$ building-block dualities --- Aharony and BBP --- and the main objective, $KP_\b^+$ duality.

\medskip\noindent{\bf Aharony duality}. The first building-block duality is the Aharony duality~\cite{Aharony:1997gp}; an IR duality between two 3d $\CN=2$ SQCDs with unitary gauge groups and $F$ flavors $(Q,\WQ)/(q,\Wq)$ that form $SU(F)_t \times SU(F)_b$ flavor symmetry. This can be represented as quiver diagrams:
\begin{equation}
\begin{tikzpicture}[thick,scale=0.8, every node/.style={scale=0.8}]%[shorten >=1pt]
  \tikzset{vertex/.style={circle,fill=white!25,minimum size=12pt,inner sep=2pt}}
  \tikzset{every loop/.style={}}%removes arrow head from all loops.
    \node[vertex] (eNc) at (-1,0) [shape=circle,draw=black,minimum size=2em] {$N$};
    \node[vertex] (eNf1) at (1,0.7) [shape=rectangle,draw=black,minimum height=2em, minimum width=2em] {$F$};
    \node[vertex] (eNf2) at (1,-0.7) [shape=rectangle,draw=black,minimum height=2em, minimum width=2em] {$F$};

    \draw[->-=.5] ([yshift= 2pt] eNc.east) to ([yshift= 0pt] eNf1.west);
    \draw[->-=.5] ([yshift= 0pt] eNf2.west) to ([yshift= -2pt] eNc.east);

    \node at (0,0.7) {\tiny$Q$};
    \node at (0,-0.7) {\tiny$\WQ$};
    \node at (0,-1.5) {$W_e=  0$};

    \node at (3.5,0.2) {$\overset{\CA}{\longleftrightarrow}$};

    \node[vertex] (mNc) at (6,0) [shape=circle,draw=black,minimum size=2em] {$N'$};
    \node[vertex] (mNf1) at (8,0.7) [shape=rectangle,draw=black,minimum height=2em, minimum width=2em] {$F$};
    \node[vertex] (mNf2) at (8,-0.7) [shape=rectangle,draw=black,minimum height=2em, minimum width=2em] {$F$};

    \draw[->-=.5] ([yshift= 0pt] mNf1.west) to ([yshift= 2pt] mNc.east);
    \draw[->-=.5] ([yshift= -2pt] mNc.east) to ([yshift= 0pt] mNf2.west);
    \draw[->-=.5] ( mNf2.north) to ( mNf1.south);

    \node at (7,0.7) {\tiny$\Wq$};
    \node at (7,-0.7) {\tiny$q$};
    \node at (8.5,0) {\tiny$M$};
    \node at (7,-1.5) {$W_m = M\, \Wq q + V^+ \hat{v}^+ + V^- \hat{v}^-$};

\end{tikzpicture}
\label{eq: Aharony quiver}
\end{equation}
with dual gauge rank $N'= F- N$. The chiral ring generators are mapped as
\begin{align}
    \WQ Q \quad &\longleftrightarrow \quad M
    \nonumber\\
    \hat V^\pm \quad &\longleftrightarrow \quad V^\pm
\end{align}
where $M$ are mesonic singlets, $\hat V^\pm$ are electric monopoles of charges $\pm1$, and $V^\pm$ are singlets coupled to the magnetic monopoles $\hat v^\pm$. Hereafter, we denote the Aharony duality by $\CA$ for the partition function derivation.

\medskip\noindent{\bf BBP duality}.
The second building-block duality is the BBP duality~\cite{Benini:2017dud} --- a monopole superpotential deformed version of the Aharony duality:
\begin{equation}
\begin{tikzpicture}[thick,scale=0.8, every node/.style={scale=0.8}]%[shorten >=1pt]
  \tikzset{vertex/.style={circle,fill=white!25,minimum size=12pt,inner sep=2pt}}
  \tikzset{every loop/.style={}}%removes arrow head from all loops.
    \node[vertex] (eNc) at (-1,0) [shape=circle,draw=black,minimum size=2em] {$N$};
    \node[vertex] (eNf1) at (1,0.7) [shape=rectangle,draw=black,minimum height=2em, minimum width=2em] {$F$};
    \node[vertex] (eNf2) at (1,-0.7) [shape=rectangle,draw=black,minimum height=2em, minimum width=2em] {$F$};

    \draw[->-=.5] ([yshift= 2pt] eNc.east) to ([yshift= 0pt] eNf1.west);
    \draw[->-=.5] ([yshift= 0pt] eNf2.west) to ([yshift= -2pt] eNc.east);

    \node at (0,0.7) {\tiny$Q$};
    \node at (0,-0.7) {\tiny$\WQ$};
    \node at (0,-1.5) {$W_e=  \hat V^+ $};

    \node at (3.5,0.2) {$\overset{\CB}{\longleftrightarrow}$};

    \node[vertex] (mNc) at (6,0) [shape=circle,draw=black,minimum size=2em] {$N''$};
    \node[vertex] (mNf1) at (8,0.7) [shape=rectangle,draw=black,minimum height=2em, minimum width=2em] {$F$};
    \node[vertex] (mNf2) at (8,-0.7) [shape=rectangle,draw=black,minimum height=2em, minimum width=2em] {$F$};

    \draw[->-=.5] ([yshift= 0pt] mNf1.west) to ([yshift= 2pt] mNc.east);
    \draw[->-=.5] ([yshift= -2pt] mNc.east) to ([yshift= 0pt] mNf2.west);
    \draw[->-=.5] ( mNf2.north) to ( mNf1.south);

    \node at (7,0.7) {\tiny$\Wq$};
    \node at (7,-0.7) {\tiny$q$};
    \node at (8.5,0) {\tiny$M$};
    \node at (7,-1.5) {$W_m = \hat v^+ + M\, \Wq q + V^- \hat{v}^-$};

\end{tikzpicture}
\label{eq: BBP quiver}
\end{equation}
The dual gauge rank is reduced by one compared to that in the Aharony duality i.e., $N'' = F-N-1$. The chiral ring generators are mapped as
\begin{align}
    \WQ Q \quad &\longleftrightarrow \quad M
    \nonumber\\
    \hat V^+ \quad &\longleftrightarrow \quad \hat v^+
    \nonumber\\
    \hat V^- \quad &\longleftrightarrow \quad  V^-
\end{align}
where the electric monopole operator appearing in the superpotential is directly mapped to its magnetic counterpart. Hereafter, we denote this BBP duality by $\CB$.

\medskip\noindent{\bf Monopole-deformed Kim--Park duality}.
\begin{table}[tbp]
\centering
\begin{tabular}{|c|c|c|c|c|}
\hline
 & $U(1)_R$ & $SU(F)_t$ & $SU(F)_b$ & $U(1)_D$ \\
\hline 
$Q$ & $r$ & $\overline{\mathbf{F}}$ & $\mathbf 1$ & $1$ \\
$\WQ$ & $r$ & $\mathbf 1$ & $\mathbf{F}$ & $1$ \\
$X$ & $2/(p+1)$ & $\mathbf 1$ & $\mathbf 1$ & $0$ \\
$\hat{V}_i^-$ & $2 F (1-r)-\frac{2}{p+1}(2 N + p - \b - 1 - i)$ & $\mathbf 1$ & $\mathbf 1$ & $-2 F$ \\
\hline
$q$ & $\frac{2}{p+1}-r$ & $\mathbf 1$ & $\overline{\mathbf{F}}$ & $-1$ \\
$\Wq$ & $\frac{2}{p+1}-r$ & $\mathbf{F}$ & $\mathbf 1$ & $-1$ \\
$x$ & $2/(p+1)$ & $\mathbf 1$ & $\mathbf 1$ & $0$ \\
$M_i$ & $2 r+2 i/(p+1)$ & $\overline{\mathbf{F}}$ & $\mathbf{F}$ & $2$ \\
$V_i^-$ & $2 F (1-r)-\frac{2}{p+1}(2 N + p - \b - 1 - i)$ & $\mathbf 1$ & $\mathbf 1$ & $-2 F$ \\
\hline
\end{tabular}
\caption{\label{tab: one monopole KP charge} Charge table of $KP_\b^+$ duality. The subscript $i$ runs from $0$ to $p-1$. $U(1)_D$ is the unbroken diagonal of $U(1)_A\times U(1)_T$; axial and topological symmetries respectively.}
\end{table}
The main focus of this paper is the $KP_\b^+$ duality~\cite{Jia:2025koz}, which can be represented by the following quiver diagram ($\b=0,\ldots,2p$):
\begin{equation}
\begin{tikzpicture}[thick,scale=0.8, every node/.style={scale=0.8}]%[shorten >=1pt]
  \tikzset{vertex/.style={circle,fill=white!25,minimum size=12pt,inner sep=2pt}}
  \tikzset{every loop/.style={}}%removes arrow head from all loops.
    \node[vertex] (eNc) at (-1,0) [shape=circle,draw=black,minimum size=2em] {$N$};
    \node[vertex] (eNf1) at (1,0.7) [shape=rectangle,draw=black,minimum height=2em, minimum width=2em] {$F$};
    \node[vertex] (eNf2) at (1,-0.7) [shape=rectangle,draw=black,minimum height=2em, minimum width=2em] {$F$};

    \draw[-to, min distance=1cm]  (eNc) edge [out=150, in=210] node {} (eNc);
    \draw[->-=.5] ([yshift= 2pt] eNc.east) to ([yshift= 0pt] eNf1.west);
    \draw[->-=.5] ([yshift= 0pt] eNf2.west) to ([yshift= -2pt] eNc.east);

    \node at (-1.6,0.6) {\tiny$X$};
    \node at (0,0.7) {\tiny$Q$};
    \node at (0,-0.7) {\tiny$\WQ$};
    \node at (0,-1.5) {$W_e= \hat V_\b^+ +  \Tr X^{p+1}$};

    \node at (3.5,0.4) {\scriptsize $KP_\b^+$};
    \node at (3.5,0) {$\longleftrightarrow$};

    \node[vertex] (mNc) at (7,0) [shape=circle,draw=black,minimum size=2em] {$\WN$};
    \node[vertex] (mNf1) at (9,0.7) [shape=rectangle,draw=black,minimum height=2em, minimum width=2em] {$F$};
    \node[vertex] (mNf2) at (9,-0.7) [shape=rectangle,draw=black,minimum height=2em, minimum width=2em] {$F$};

    \draw[-to, min distance=1cm]  (mNc) edge [out=150, in=210] node {} (mNc);
    \draw[->-=.5] ([yshift= 0pt] mNf1.west) to ([yshift= 2pt] mNc.east);
    \draw[->-=.5] ([yshift= -2pt] mNc.east) to ([yshift= 0pt] mNf2.west);
    \draw[->-=.5] ( mNf2.north) to ( mNf1.south);

    \node at (6.4,0.6) {\scriptsize$x$};
    \node at (8,0.7) {\tiny$\Wq$};
    \node at (8,-0.7) {\tiny$q$};
    \node at (9.5,0) {\tiny$M_i$};
    \node at (6,-1.5) {$W_m = \hat v_\b^+ + \Tr x^{p+1}$};
    \node at (9,-2) {$\!+\! 
    \sum_{i=0}^{p-1} M_i \,\Wq x^{p\texttt{-}i\texttt{-}1} q 
    +
    \sum_{i=0}^{p-1}
    V_i^- \hat v_{p\texttt{-}i\texttt{-}1}^-$};

\end{tikzpicture}
\label{eq: KPb quiver}
\end{equation}
For $\b \geq p$, $\hat V_\b^\pm$ denotes a composite of dressed monopoles and adjoint traces, following~\cite{Jia:2025koz}. The charges of matter fields are summarized in Table \ref{tab: one monopole KP charge}. The dual gauge rank is $(\WN \geq 0)$
\begin{align}
    \WN = pF-N-p+\b,
\end{align}
and the chiral ring generators are mapped as $(i=0,\cdots,p-1)$
\begin{align}
    \Tr X^i
    \qquad&\longleftrightarrow\qquad
    \Tr x^i
    \nonumber\\
    \WQ X^i Q
    \qquad&\longleftrightarrow\qquad
    M_i
    \nonumber\\
    \hat V_{i}^+
    \qquad&\longleftrightarrow\qquad
    \hat v_{i}^+
    \nonumber\\
    \hat V_i^-
    \qquad&\longleftrightarrow\qquad
    V_i^-
    \,.
\end{align}
The $KP_\b^-$ duality follows by simply flipping all magnetic signs. For $p=1$ and $\b=0$, the adjoint fields become massive, and the $KP_0^+$ duality reduces to the BBP duality.

\subsection{S-confinement of \texorpdfstring{$\CC_p[U(N)]$}{Cp[U(N)]}}
For any positive integers $p$ and $N$, the 3d $\CN=2$ linear quiver gauge theory $\CC_p[U(N)]$ introduced in~\cite{Benvenuti:2024glr} is constructed as follows:
\begin{equation}
\begin{tikzpicture}[thick,scale=0.8, every node/.style={scale=0.8}]%[shorten >=1pt]
  \tikzset{vertex/.style={circle,fill=white!25,minimum size=12pt,inner sep=2pt}}
  \tikzset{every loop/.style={}}%removes arrow head from all loops.
    \node at (-2,0) {$\CC_p[U(N)]\quad:\quad$ };    

    \node[vertex] (A1) at (0,0) [shape=circle,draw=black,minimum size=2em] {$m_1$};
    \node[vertex] (A2) at (1.5,0) [shape=circle,draw=black,minimum size=2em] {$m_2$};
    \node[vertex] (Acd) at (3,0)  {$\cdots$};
    \node[vertex] (Ap) at (4.5,0) [shape=circle,draw=black,minimum size=2em] { \scriptsize $m_{p\texttt{-}1}\!\!$};
    \node[vertex] (AF) at (6,0) [shape=rectangle,draw=black,minimum height=2em, minimum width=2em] {$N$};

    \draw[-to, min distance=0.8cm]  (A1) edge [out=120, in=60] node {} (A1);
    \draw[-to, min distance=0.8cm]  (A2) edge [out=120, in=60] node {} (A2);
    \draw[-to, min distance=0.8cm]  (Ap) edge [out=120, in=60] node {} (Ap);
    \draw[->-=.5] ([yshift= 2pt] A1.east) to ([yshift= 2pt] A2.west);
    \draw[->-=.5] ([yshift= -2pt] A2.west) to ([yshift= -2pt] A1.east);
    \draw[->-=.5] ([yshift= 2pt] A2.east) to ([yshift= 2pt] Acd.west);
    \draw[->-=.5] ([yshift= -2pt] Acd.west) to ([yshift= -2pt] A2.east);
    \draw[->-=.5] ([yshift= 2pt] Acd.east) to ([yshift= 2pt] Ap.west);
    \draw[->-=.5] ([yshift= -2pt] Ap.west) to ([yshift= -2pt] Acd.east);
    \draw[->-=.5] ([yshift= 2pt] Ap.east) to ([yshift= 2pt] AF.west);
    \draw[->-=.5] ([yshift= -2pt] AF.west) to ([yshift= -2pt] Ap.east);

\end{tikzpicture}
\label{eq: Cp quiver}
\end{equation}
The ranks are given by
\begin{align}
    m_j = j n + \a_j
    \;\;,\quad
    j=1,\cdots,p
    \,,
\end{align}
with $n = \lfloor N / p \rfloor$ and $\vec{\a} = (\overbrace{0,\cdots,0}^{p-\a},1,2,\cdots,\a)$ for $\a:=N \!\!\mod p$. Thus, $m_p = pn+\a= N$ by construction. The superpotential of this theory is given by~\footnote{Unlike $\CC_p[SU(N)]$ in~\cite{Benvenuti:2024glr}, $\CC_p[U(N)]$ has no flipping field for $\hat{V}^{(1,p-1),-}$ at $\a=0$, as we identify this operator with a component of the confined $U(N)$-adjoint rather than the $SU(N)$-adjoint. Also, the superpotential assumes $N\geq p$; for $N<p$, we use $\CC_p[U(N)]$ only as the
partition function block \eqref{eq: small N}, with zero-rank nodes omitted.}
\begin{align}
    W
    =
    W_{\CN=4} 
    + 
    \sum_{i=1}^{p-1} \hat V^{(i),+}
    +
    \Tr( \phi^{\d_{ 0 \a}}
    \hat V^{(1,p-1),-})
    \,,
    \label{eq: CpN superpotential}
\end{align}
where $W_{\CN=4}$ denotes the $\CN=4$-like cubic terms, $\hat V^{(i),+}$ is a bare monopole operator with magnetic charge $+\d_{ij}$ under $U(m_j)$, and $\hat V^{(1,p-1),-}$ carries magnetic charge $-1$ under each gauge group. For $\a=0$, the third term becomes a monopole operator dressed by the adjoint chiral field $\phi$.

\medskip\noindent{\bf S-confinement}.
The $\CC_p[U(N)]$ theory is known to be dual to an adjoint chiral multiplet $X$ of the $U(N)$ flavor symmetry, together with singlets $\eta_i$ :
\begin{equation}
    \begin{tikzpicture}[thick,scale=0.8, every node/.style={scale=0.8}]%[shorten >=1pt]
      \tikzset{vertex/.style={circle,fill=white!25,minimum size=12pt,inner sep=2pt}}
      \tikzset{every loop/.style={}}%removes arrow head from all loops.
        \node[vertex] (N) at (0,0) [shape=rectangle,draw=black,minimum height=2em, minimum width=2em] {$N$};
    
        \draw[-to, min distance=1cm]  (N) edge [out=120, in=60] node {} (N);
    
        \node at (-4,0) {$\CC_p[U(N)]$};
        \node at (-2.2,0.2) {$\overset{\text{Dual}}{\longleftrightarrow}$};
        
        \node at (0,1.2) {\tiny $X$};
        \node at (1.7,0.2) {\small$\quad W=\Tr X^{p+1}$};
        \node at (3.2,-0.2) {\small$+\sum_{i=1}^{\min(\a,p\texttt{-}\a)} \eta_i \Tr X^i$};
    \end{tikzpicture}
    \label{eq: confinement}
\end{equation}
for which we provide an alternative partition function derivation assuming only Aharony and BBP dualities in Appendix~\ref{app: conf derivation}.~\footnote{The identity was already proven in~\cite{Benvenuti:2024glr}.} The duality \eqref{eq: confinement} allows us to replace the adjoint chiral multiplet with a $\CC_p[U(N)]$ quiver tail, a procedure known as {\it deconfinement}~\cite{Berkooz:1995km,Pouliot:1995me,Luty:1996cg}. Using this, the Kim--Park duality~\cite{Kim:2013cma} can be derived by deconfining the adjoint, applying a sequence of node-wise Aharony dualities, and finally confining the resulting quiver back to the dual adjoint SQCD~\cite{Benvenuti:2024glr,Hwang:2024hhy}. To ensure a consistent Coulomb branch spectrum, an appropriate BF term is required during deconfinement, which is encoded by shifting the FI-parameter $\t$ by
\begin{align}
    \t_p = \t - [N-1]_p\D
    \,,
\end{align}
where $[l]_p:=l\!\!\mod p$, $\D \equiv \frac{1}{p+1}$, and $\t_p$ is the shifted FI-parameter after the deconfinement.

\section{Partition function derivation}
In this section, we establish the matching of the ellipsoid partition functions for the 3d $\CN=2$ $KP_\b^+$ dual pair. We begin by introducing a diagrammatic quiver representation of the partition function. 
The derivation proceeds as follows. We first deconfine the electric theory into a $\CC_p[U(N)]$ linear quiver tail.
Applying the two elementary dualities $\CA$ and $\CB$ node by node then produces a new tail, which confines back.
Finally, using the rank-reducing duality chain, we arrive at the expected $KP_\b^+$ magnetic theory.

\subsection{Quiver expression of the ellipsoid partition function}
The ellipsoid partition function of a 3d $\CN=2$ gauge theory with gauge group $G$ and chiral multiplets $\Phi$ of gauge weight $\r_\Phi$, flavor weight $f_\Phi$, and R-charge $r_\Phi$ reads~\cite{Hama:2011ea}
\begin{align}
    \CZ_{S_b^3}(m,\xi) = \frac{1}{|W_G|} 
    \int \prod_{i=1}^{\text{rank}(G)} d \s_i
    e^{2\pi i \xi \Tr \s}
    \frac{
    \prod_{\Phi}\prod_{\r_{\Phi},f_{\Phi}} 
    s_b \big((1-r_\Phi)c_b - \rho_\Phi(\s) - f_\Phi(m) \big)
    }
    {
    \prod_{\a} s_b\big(\a(\s) + c_b\big)
    }
    \label{eq: S_b^3 ptf integral formula}
\end{align}
where $c_b:=i(b+b^{-1})/2$, $W_G$ denotes the Weyl group of $G$, $\a$ is a root of $G$, $\xi$ is the FI-parameter, $\s$ parametrizes the Coulomb branch, and $m$ is the real mass parameter associated with the flavor symmetry. The double-sine function $s_b(\s)$ is defined by
\begin{align}
    s_b(\s) := \prod_{l,k\geq 0}
    \frac{l b + k b^{-1} -i c_b - i \s}{l b + k b^{-1} -i c_b + i \s} \,,
\end{align}
which represents the contribution of a chiral multiplet. Note that $s_b(\s)s_b(-\s)=1$, reflecting the cancellation of two massive chiral multiplets paired by a superpotential mass term.~\footnote{
The partition function identities below hold by meromorphic continuation in complexified masses and FI-parameters~\cite{Willett:2011gp}, with reflection pairs canceled via $s_b(z)s_b(-z)=1$ {\it before} taking singular limits.
}

We use a quiver representation of the ellipsoid partition function~\cite{Hwang:2024hhy,Benvenuti:2026xcv}, focusing on unitary gauge and flavor groups with bi-fundamental or adjoint chirals. The quiver components are defined as follows:
\begin{equation}
    \begin{tikzpicture}[thick,baseline={([yshift=-.5ex]current bounding box.center)}]
  \tikzset{vertex/.style={circle,fill=white!25,minimum size=12pt,inner sep=2pt}}
  \tikzset{every loop/.style={}}%removes arrow head from all loops.
    \node[vertex] (N) at (0,0) [shape=circle,draw=black,minimum size=2em] {$N$};
    \node at (0,-0.6) {\scriptsize $\t$};
    \end{tikzpicture}
    \;\;\Longleftrightarrow \;\;
    \frac{1}{N!} \int \prod_{i=1}^{N} d\s_i \; e^{2\pi i c_b \t \sum_{i=1}^{N} \s_i} \frac{1}{\prod_{i\neq j}^{N} s_b ( \s_i - \s_j +c_b)} \,,
    \label{eq: circular node}
\end{equation}
\begin{equation}
\begin{tikzpicture}[baseline={([yshift=-.5ex]current bounding box.center)}]
  \tikzset{vertex/.style={circle,fill=white!25,minimum size=12pt,inner sep=2pt}}
  \tikzset{every loop/.style={}}%removes arrow head from all loops.
    \node[vertex] (N) at (0,0) {$N$};
    \node[vertex] (M) at (1.5,0) {$M$};
    \node at (0.7,0.3) {$r$};
    \draw[->-=.55] ([yshift= 0pt] N.east) to ([yshift= 0pt] M.west);
    \end{tikzpicture}
    \;\;\Longleftrightarrow\;\;
    \prod_{i=1}^{N} \prod_{j=1}^{M} 
    s_b \Big( y_j - x_i + c_b(1-r) \Big) \,,
    \qquad\qquad\qquad\quad
    \label{eq: arrow}
\end{equation}
\begin{equation}
    +\big\{r\big\}
    \;\;\Longleftrightarrow\;\;
    s_b \Big((1-r)c_b \Big) \,.
    \qquad\qquad\qquad\qquad\qquad\qquad
    \label{eq: singlet}
\end{equation}
The circular node \eqref{eq: circular node} represents a $U(N)$ gauge group with its normalized FI-parameter $\t$ displayed below. The arrow in \eqref{eq: arrow} represents a bi-fundamental chiral multiplet of $U(N)$ and $U(M)$, with fugacities $x_i$ and $y_j$, respectively, while a self-loop represents an adjoint chiral multiplet. The R-charge $r$ is indicated next to each arrow. Singlets, when present, are grouped in curly brackets with their R-charges, as in \eqref{eq: singlet}. We now present a few examples relevant to the following discussion:
\begin{itemize}
    \item \medskip\noindent{\bf Basic dualities}.
The dualities $\CA$ and $\CB$ in a linear quiver can be expressed as
\begin{equation}
\begin{tikzpicture}[thick,scale=0.9, every node/.style={scale=0.8}]%[shorten >=1pt]
  \tikzset{vertex/.style={circle,fill=white!25,minimum size=12pt,inner sep=2pt}}
  \tikzset{every loop/.style={}}%removes arrow head from all loops.
  % Original theory
    \node[vertex] (n1) at (0-1.5,0) [shape=circle,draw=black,minimum size=2em] {$M$};
    \node[vertex] (n2) at (1.5-1.5,0) [shape=circle,draw=black,minimum height=2em, minimum width=2em] {$N$};
    \node[vertex] (n3) at (3-1.5,0) [shape=circle,draw=black,minimum height=2em, minimum width=2em] {$L$};

    \draw[->-=.5] ([yshift= 2pt] n1.east) to ([yshift= 2pt] n2.west);
    \draw[->-=.5] ([yshift= -2pt] n2.west) to ([yshift= -2pt] n1.east);
    \draw[->-=.5] ([yshift= 2pt] n2.east) to ([yshift= 2pt] n3.west);
    \draw[->-=.5] ([yshift= -2pt] n3.west) to ([yshift= -2pt] n2.east);

    \node at (-0.7-1.5,0) {$\cdots$};
    \node at (3.7-1.5,0) {$\cdots$};
    \node at (0.75-1.5,0.4) {\scriptsize $r_1$};
    \node at (2.25-1.5,0.4) {\scriptsize $r_2$};
    \node at (0-1.5,-0.5) {\scriptsize $\th$};
    \node at (1.5-1.5,-0.5) {\scriptsize $\t$};
    \node at (3-1.5,-0.5) {\scriptsize $\l$};
    
    % Aharony dual
    \node[vertex] (N1) at (0+5,0+1) [shape=circle,draw=black,minimum size=2em] {$M$};
    \node[vertex] (N2) at (1.5+5,0+1) [shape=circle,draw=black,minimum height=2em, minimum width=2em] {$N'$};
    \node[vertex] (N3) at (3+5,0+1) [shape=circle,draw=black,minimum height=2em, minimum width=2em] {$L$};

    \draw[-to, min distance=1cm]  (N1) edge [out=120, in=60] node {} (N1);
    \draw[-to, min distance=1cm]  (N3) edge [out=120, in=60] node {} (N3);
    \draw[->-=.5] ([yshift= 2pt] N1.east) to ([yshift= 2pt] N2.west);
    \draw[->-=.5] ([yshift= -2pt] N2.west) to ([yshift= -2pt] N1.east);
    \draw[->-=.5] ([yshift= 2pt] N2.east) to ([yshift= 2pt] N3.west);
    \draw[->-=.5] ([yshift= -2pt] N3.west) to ([yshift= -2pt] N2.east);
    \draw[->-=.5] ([xshift= 6pt,yshift= -2pt] N1.north) to [bend right=-60] ([xshift= -6pt,yshift= -2pt] N3.north);
    \draw[->-=.5] ([xshift= -9pt,yshift= -4pt] N3.north) to [bend right=60] ([xshift= 9pt,yshift= -4pt] N1.north);

    \node at (-0.7+5,0+1) {$\cdots$};
    \node at (3.7+5,0+1) {$\cdots$};
    \node at (0+5,1.2+1) {\scriptsize $2r_1$};
    \node at (3+5,1.2+1) {\scriptsize $2r_2$};
    \node at (0.75+5,0.25+1) {\scriptsize $1\texttt{-}r_1$};
    \node at (2.25+5,0.25+1) {\scriptsize $1\texttt{-}r_2$};
    \node at (1.5+5,1.2+1) {\scriptsize $r_1 \texttt{+} r_2$};
    \node at (0+5,-0.7+1.2) {\scriptsize $\th+\t$};
    \node at (1.5+5,-0.7+1.2) {\scriptsize $-\t$};
    \node at (3+5,-0.7+1.2) {\scriptsize $\l+\t$};

    \node at (6.5,0.7-0.5) {\scriptsize $+\{M(1\!-\!r_1) \!+\! L (1\!-\!r_2) \!-\!N \!+\! 1 \pm \t\}$};

    % BBP1 dual
    \node[vertex] (b1) at (5,0-3+1) [shape=circle,draw=black,minimum size=2em] {$M$};
    \node[vertex] (b2) at (6.5,0-3+1) [shape=circle,draw=black,minimum height=2em, minimum width=2em] {$N''$};
    \node[vertex] (b3) at (8,0-3+1) [shape=circle,draw=black,minimum height=2em, minimum width=2em] {$L$};

    \draw[-to, min distance=1cm]  (b1) edge [out=120, in=60] node {} (b1);
    \draw[-to, min distance=1cm]  (b3) edge [out=120, in=60] node {} (b3);
    \draw[->-=.5] ([yshift= 2pt] b1.east) to ([yshift= 2pt] b2.west);
    \draw[->-=.5] ([yshift= -2pt] b2.west) to ([yshift= -2pt] b1.east);
    \draw[->-=.5] ([yshift= 2pt] b2.east) to ([yshift= 2pt] b3.west);
    \draw[->-=.5] ([yshift= -2pt] b3.west) to ([yshift= -2pt] b2.east);
    \draw[->-=.5] ([xshift= 6pt,yshift= -2pt] b1.north) to [bend right=-60] ([xshift= -6pt,yshift= -2pt] b3.north);
    \draw[->-=.5] ([xshift= -9pt,yshift= -4pt] b3.north) to [bend right=60] ([xshift= 9pt,yshift= -4pt] b1.north);

    \node at (4.3,0-3+1) {$\cdots$};
    \node at (8.7,0-3+1) {$\cdots$};
    \node at (5,1.2-3+1) {\scriptsize $2r_1$};
    \node at (8,1.2-3+1) {\scriptsize $2r_2$};
    \node at (5.75,0.25-3+1) {\scriptsize $1\texttt{-}r_1$};
    \node at (7.25,0.25-3+1) {\scriptsize $1\texttt{-}r_2$};
    \node at (1.5+5,1.2-3+1) {\scriptsize $r_1 \texttt{+} r_2$};
    \node at (5,-0.7-3+1.2) {\scriptsize $\th\!+\!\t \!-\! r_1$};
    \node at (6.5,-0.7-3+1.2) {\scriptsize $\!-\!\t \!+\! 1$};
    \node at (8,-0.7-3+1.2) {\scriptsize $\l\!+\!\t \!-\! r_2$};

    \node at (1.4+5,1.7-3+0.8) {\scriptsize $+\{2- 2\t\}$};

    % Arrows
    
    \node at (3,1-0.4) 
    {\begin{tikzpicture}
      \node [rotate=45] {\textcolor{black}{$=\!\!=$}};    
    \end{tikzpicture}
    };
    \node at (2.8,1.3-0.4) {\scriptsize $\CA$};
    
    \node at (3,-1.5+0.4) 
    {\begin{tikzpicture}
      \node [rotate=135] {\textcolor{black}{$=\!\!=$}};    
    \end{tikzpicture}
    };
    \node at (3.1,-1.2+0.4) {\scriptsize $\CB$};
    \node at (2.9,-2+0.4) {\scriptsize if \eqref{eq: B cond}};

\end{tikzpicture}
\label{eq: basic duality in quiver}
\end{equation}
with $N' = M+L-N$ and $N''=M+L-N-1$. Importantly, $\CB$ is applicable only when the monopole superpotential term has R-charge $2$.
\begin{align}
    M(1-r_1) + L(1-r_2) - N + 1 + \t = 2
    \,.
    \label{eq: B cond}
\end{align}

\item \medskip\noindent{\bf The confinement}. 
The s-confinement of $\CC_p[U(N)]$ in \eqref{eq: confinement} can also be expressed as follows (see Appendix~\ref{app: conf derivation} for a derivation):
\begin{equation}
\begin{tikzpicture}[thick,scale=1.1, every node/.style={scale=0.8}]%[shorten >=1pt]
\tikzset{vertex/.style={circle,fill=white!25,minimum size=12pt,inner sep=2pt}}
  \tikzset{every loop/.style={}}%removes arrow head from all loops.
    % Theory A
    \node[vertex] (A1) at (0,0) [shape=circle,draw=black,minimum size=2em] {$m_1$};
    \node[vertex] (A2) at (1.2,0) [shape=circle,draw=black,minimum size=2em] {$m_2$};
    \node[vertex] (Acd) at (2.4,0)  {$\cdots$};
    \node[vertex] (Ap) at (3.6,0) [shape=circle,draw=black,minimum size=2em] { \scriptsize $m_{p\texttt{-}1}\!\!$};
    \node[vertex] (AF) at (4.8,0) [shape=rectangle,draw=black,minimum height=2em, minimum width=2em] {$N$};

    \draw[-to, min distance=0.8cm]  (A1) edge [out=120, in=60] node {} (A1);
    \draw[-to, min distance=0.8cm]  (A2) edge [out=120, in=60] node {} (A2);
    \draw[-to, min distance=0.8cm]  (Ap) edge [out=120, in=60] node {} (Ap);
    \draw[->-=.5] ([yshift= 2pt] A1.east) to ([yshift= 2pt] A2.west);
    \draw[->-=.5] ([yshift= -2pt] A2.west) to ([yshift= -2pt] A1.east);
    \draw[->-=.5] ([yshift= 2pt] A2.east) to ([yshift= 2pt] Acd.west);
    \draw[->-=.5] ([yshift= -2pt] Acd.west) to ([yshift= -2pt] A2.east);
    \draw[->-=.5] ([yshift= 2pt] Acd.east) to ([yshift= 2pt] Ap.west);
    \draw[->-=.5] ([yshift= -2pt] Ap.west) to ([yshift= -2pt] Acd.east);
    \draw[->-=.5] ([yshift= 2pt] Ap.east) to ([yshift= 2pt] AF.west);
    \draw[->-=.5] ([yshift= -2pt] AF.west) to ([yshift= -2pt] Ap.east);

    \node at (0.6,0.4) {\scriptsize $\D$};
    \node at (1.8,0.4) {\scriptsize $\D$};
    \node at (3,0.4) {\scriptsize $\D$};
    \node at (4.2,0.4) {\scriptsize $\D$};
    \node at (0,1.1) {\scriptsize $2\texttt{-}2\D$};
    \node at (1.2,1.1) {\scriptsize $2\texttt{-}2\D$};
    \node at (3.6,1.1) {\scriptsize $2\texttt{-}2\D$};
    \node at (5.2,1) {$+\big\{2\D j\big\}_{j=1}^{\min(\a,p-\a)}$};

    \node at (0,-0.5) {\scriptsize $\t_1$};
    \node at (1.2,-0.5) {\scriptsize $\t_2$};
    \node at (3.6,-0.5) {\scriptsize $\t_{p-1}$};

    % Theory F
    \node[vertex] (FF) at (7.5,0) [shape=rectangle,draw=black,minimum height=2em, minimum width=2em] {$N$};
    \draw[-to, min distance=0.8cm]  (FF) edge [out=120, in=60] node {} (FF);

    \node at (7.5,1.1) {\scriptsize $2\D$};

    % Arrows

    \node at (6.2,0) 
    {\begin{tikzpicture}
      \node [rotate=0] {\Large\textcolor{black}{$=\!\!=$}};    
    \end{tikzpicture}
    };
    
\end{tikzpicture}
\end{equation}
The singlet contributions from $\eta_j$ have been moved to the left-hand side. The FI-parameters $\t_j$ are fixed by the superpotential \eqref{eq: CpN superpotential} as
\begin{align}
    \t_j = (2-\d_{j,p-\a}\,p)\D
    \,.
\end{align}

\item \medskip\noindent{\bf Rank-reducing}. We also employ the rank-reducing duality chain of~\cite{Hwang:2024hhy}; see Appendix~\ref{app: rr} for a partition function derivation. Each iteration reduces the gauge rank by one at a specific FI-parameter, producing a pair of singlets. After $N$ iterations, the gauge node disappears, leaving $2N$ singlets:
\begin{equation}
    \begin{tikzpicture}[thick,scale=0.9, every node/.style={scale=0.8}]%[shorten >=1pt]
    \tikzset{vertex/.style={circle,fill=white!25,minimum size=12pt,inner sep=2pt}}
      \tikzset{every loop/.style={}}%removes arrow head from all loops.
        % Theory A
        \node[vertex] (AN) at (0,0) [shape=circle,draw=black,minimum size=2em] {$N$};
        \node[vertex] (AF) at (1.5,0) [shape=circle,draw=black,minimum height=2em, minimum width=2em] {$M$};
        \node[vertex] (DM) at (4.2,0) [shape=circle,draw=black,minimum height=2em, minimum width=2em] {$M$};

        \node at (2.2,0) {$\cdots$};
        \node at (4.9,0) {$\cdots$};
    
        \draw[-to, min distance=1cm]  (AN) edge [out=120, in=60] node {} (AN);
        \draw[->-=.5] ([yshift= 2pt] AN.east) to ([yshift= 2pt] AF.west);
        \draw[->-=.5] ([yshift= -2pt] AF.west) to ([yshift= -2pt] AN.east);
    
        \node at (0,1.2) {\scriptsize $2\d$};
        \node at (0.75,0.4) {\scriptsize $r$};
        \node at (0,-0.6) {\scriptsize $\pm M(1\texttt{-}r)$};
        \node at (1.5,-0.6) {\scriptsize $\k$};
        \node at (4.2,-0.6) {\scriptsize $\k \pm N(1-r)$};

        \node at (3,0) { $=\!\!\!=$};

        \node at (4.5,0.8) {$+ \big\{\; 2\d j  \;\big\}_{j=1-N}^{N} $};
    \end{tikzpicture}
    \label{eq: rr}
\end{equation}
Note that the FI-parameter condition $\t=\pm M(1-r)$ for applying the rank-reducing duality chain is compatible with turning on a VEV for the maximally dressed monopole operator $\hat V_{\phi^{N-1}}^\mp$.

\end{itemize}
We emphasize that all these identities are derived {\it solely} from $\CA$ and $\CB$.

\subsection{Structure of the derivation}
We establish the matching of partition functions under the $KP_\b^+$ duality:

\begin{equation}
    \begin{tikzpicture}[thick]%[shorten >=1pt]
  \tikzset{vertex/.style={circle,fill=white!25,minimum size=12pt,inner sep=2pt}}
  \tikzset{every loop/.style={}}%removes arrow head from all loops.
    \node[vertex] (eNc) at (-1,0) [shape=circle,draw=black,minimum size=2em] {$N$};
    \node[vertex] (eNf1) at (1,0.7) [shape=rectangle,draw=black,minimum height=2em, minimum width=2em] {$F$};
    \node[vertex] (eNf2) at (1,-0.7) [shape=rectangle,draw=black,minimum height=2em, minimum width=2em] {$F$};

    \draw[->-=.5] ([yshift= 2pt] eNc.east) to ([yshift= 0pt] eNf1.west);
    \draw[->-=.5] ([yshift= 0pt] eNf2.west) to ([yshift= -2pt] eNc.east);
    \draw[-to, min distance=1cm]  (eNc) edge [out=120, in=60] node {} (eNc);

    \node at (-1,-0.6) {\tiny$\t$};
    \node at (-1,1.3) {\tiny$2\D$};
    \node at (0,0.7) {\tiny$r$};
    \node at (0,-0.7) {\tiny$r$};

    \node at (3.5,0.3) {\scriptsize $KP_\b^+$};
    \node at (3.5,-0.1) {$=\!\!=$};

    \node[vertex] (mNc) at (6,0) [shape=circle,draw=black,minimum size=2em] {$\WN$};
    \node[vertex] (mNf1) at (8,0.7) [shape=rectangle,draw=black,minimum height=2em, minimum width=2em] {$F$};
    \node[vertex] (mNf2) at (8,-0.7) [shape=rectangle,draw=black,minimum height=2em, minimum width=2em] {$F$};

    \draw[->-=.5] ([yshift= 0pt] mNf1.west) to ([yshift= 2pt] mNc.east);
    \draw[->-=.5] ([yshift= -2pt] mNc.east) to ([yshift= 0pt] mNf2.west);
    \draw[->-=.5] ([xshift= -5pt] mNf2.north) to ([xshift= -5pt] mNf1.south);
    \draw[->-=.5] ([xshift= 5pt] mNf2.north) to ([xshift= 5pt] mNf1.south);
    \draw[-to, min distance=1cm]  (mNc) edge [out=120, in=60] node {} (mNc);

    \node at (6,-0.6) {\tiny$\tilde{\t}$};
    \node at (6,1.3) {\tiny$2\D$};
    \node at (7,0.7) {\tiny$2\D\texttt{-}r$};
    \node at (7,-0.7) {\tiny$2\D\texttt{-}r$};
    \node at (8-0.1,0) {\tiny $\cdot$};
    \node at (8,0) {\tiny $\cdot$};
    \node at (8+0.1,0) {\tiny $\cdot$};
    \node at (9,0.1) {\tiny$2r \texttt{+} 2j\D$};
    \node at (9.3,-0.2) {\tiny$j=0,\cdots,p\texttt{-}1$};
    \node at (8,1.5) {\tiny $+\{ R_0 + 2\D j - \t \}_{j=0}^{p\texttt{-}1}$};

\end{tikzpicture}
\label{eq: the dual}
\end{equation}
with $\WN = pF-N-p+\b$. The FI-parameters are constrained by requiring the monopole superpotential terms to have R-charge $2$:
\begin{align}
    &F(1-r) - 2\D(N-1 - \b) + \t = 2
    \,,
    \nonumber\\
    &F(1-2\D + r) - 2\D (\WN - 1 - \b) + \tilde{\t} = 2
    \,,
    \label{eq: FI cond}
\end{align}
which implies $\tilde{\t}=-\t+2\D(p-\b)$. The magnetic singlet contributions, with $R_0\equiv F(1-r)-2\D(N-1)$, correspond to singlets coupled to the dual monopole operators.~\footnote{
The explicit duality sequences below are restricted to non-negative
intermediate gauge ranks. Rank-zero steps are included only when
covered by the corresponding confining partition-function identities
or their inverses.
}

\medskip
We sketch the derivation of \eqref{eq: the dual} in three cases depending on the value of $\b$:
\begin{itemize}
    \item $\b=0$ or $p$ :
    We first consider the cases with $\b=0$ and $p$; see Figure~\ref{fig: [b]=0}. We deconfine the adjoint in (e) into a $\CC_p[U(N)]$ quiver tail as in (a). Then we successively apply $\CB$ for $\b=0$ or $\CA$ for $\b=p$, starting from the $U(N)$ node and proceeding to the last node. This produces another $\CC_p[U(\WN)]$ tail in (b), whose confinement yields the expected magnetic partition function (m).

    \item $0<\b<p$ :
    Similarly, for $0<\b<p$, we deconfine the adjoint in (e) to obtain (a), as illustrated in Figure~\ref{fig: <b<}. A sequence of $\CA$ and $\CB$ dualities, detailed below, leads us to a two-node quiver (b). The $U(M)$ node is subsequently eliminated via the rank-reducing duality chain, leading to the expected magnetic partition function (m).

    \item $p<\b \leq 2p$ :
    Assuming the partition function identities for the original KP duality and $KP_\g^+$ with $0\leq\g< p$, we establish the matching for $p<\b\leq 2p$ by successively applying KP, $KP_{2p-\b}^+$, and KP dualities, as illustrated in Figure~\ref{fig: big b}. Note that $0\leq 2p-\b<p$.
    
\end{itemize}
In what follows, we verify the derivation of each case described above.

\subsection{The derivation}

\subsubsection{For \texorpdfstring{$\b=0$, or $p$}{[b]=0}}
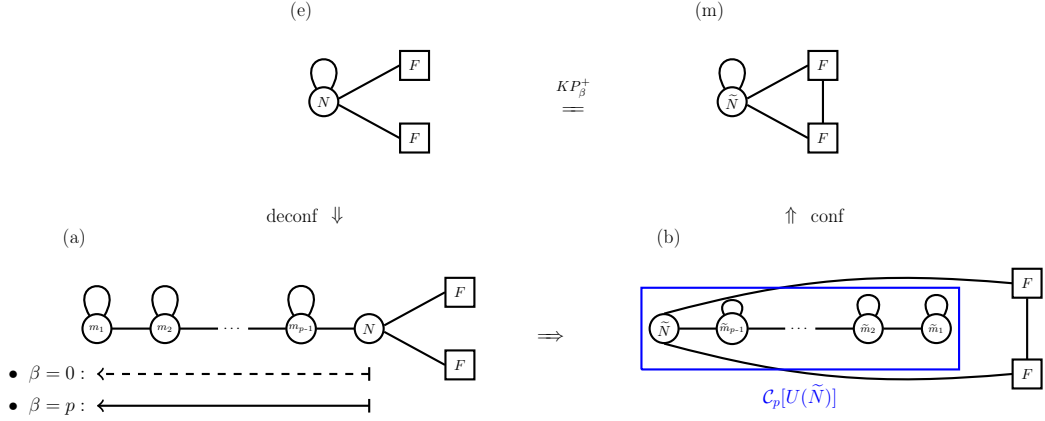
\begin{figure}[thbp]
\centering
\begin{tikzpicture}[thick,scale=0.6, every node/.style={scale=0.5}]%[shorten >=1pt]%[shorten >=1pt]
  \tikzset{vertex/.style={circle,fill=white!25,minimum size=12pt,inner sep=2pt}}
  \tikzset{every loop/.style={}}%removes arrow head from all loops.
    
    % LHS Theory
    \node[vertex] (m1) at (0,0) [shape=circle,draw=black,minimum size=2em] {\scriptsize $\!m_1\!$};
    \node[vertex] (m2) at (1.5,0) [shape=circle,draw=black,minimum size=2em] {\scriptsize $\!m_2\!$};
    \node (ecdot) at (3,0) {$\cdots$};
    \node[vertex] (mp-1) at (4.5,0) [shape=circle,draw=black,minimum size=2em] {\scriptsize $\!m_{p\texttt{-}1}\!$};
    \node[vertex] (Nc) at (6,0) [shape=circle,draw=black,minimum size=2em] {$N$};
    \node[vertex] (eNf1) at (8,0.8) [shape=rectangle,draw=black,minimum height=2em, minimum width=2em] {$F$};
    \node[vertex] (eNf2) at (8,-0.8) [shape=rectangle,draw=black,minimum height=2em, minimum width=2em] {$F$};

    \draw[-, min distance=1cm]  (m1) edge [out=120, in=60] node {} (m1);
    \draw[-, min distance=1cm]  (m2) edge [out=120, in=60] node {} (m2);
    \draw[-] (m1.east) to (m2.west);
    \draw[-] (m2.east) to (ecdot.west);
    \draw[-] (ecdot.east) to (mp-1.west);
    \draw[-, min distance=1cm]  (mp-1) edge [out=120, in=60] node {} (mp-1);
    \draw[-] (mp-1.east) to (Nc.west);
    \draw[-] ([yshift= 2pt] Nc.east) to ([yshift= 0pt] eNf1.west);
    \draw[-] ([yshift= 0pt] eNf2.west) to ([yshift= -2pt] Nc.east);

    \node at (-0.5,2) {\Large (a)};

    \node at (10,-0.2) {\Large $=\!\!\Rightarrow$};

    %Arrows
    \coordinate (an) at (6,-1+0.15);
    \coordinate (as) at (6,-1-0.15);
    \coordinate (ae) at (6,-1);
    \coordinate (aw) at (0,-1);

    \draw[-] (an) to (as);
    \draw[dashed][-to] (ae) to (aw);
    \node at (-1.1,-1) {\Large $\bullet\;\; \b = 0:$};

    \coordinate (bn) at (6,-1-0.7+0.15);
    \coordinate (bs) at (6,-1-0.7-0.15);
    \coordinate (be) at (6,-1-0.7);
    \coordinate (bw) at (0,-1-0.7);

    \draw[-] (bn) to (bs);
    \draw[-to] (be) to (bw);
    \node at (-1.1,-1.7) {\Large $\bullet \;\; \b = p:$};

    % RHS Theory
    \node[vertex] (dm1) at (12.5,0) [shape=circle,draw=black,minimum size=2em] {$\WN$};
    \node[vertex] (dm2) at (1.5+12.5,0) [shape=circle,draw=black,minimum size=2em] {\scriptsize $\!\Wm_{p\texttt{-}1}\!$};
    \node (dcdot) at (3+12.5,0) {$\cdots$};
    \node[vertex] (dmp) at (4.5+12.5,0) [shape=circle,draw=black,minimum size=2em] {\scriptsize $\!\Wm_2\!$};
    \node[vertex] (dNc) at (6+12.5,0) [shape=circle,draw=black,minimum size=2em] {\scriptsize $\!\Wm_1\!$};
    \node[vertex] (dNf1) at (8+12.5,1) [shape=rectangle,draw=black,minimum height=2em, minimum width=2em] {$F$};
    \node[vertex] (dNf2) at (8+12.5,-1) [shape=rectangle,draw=black,minimum height=2em, minimum width=2em] {$F$};

    \draw[-, min distance=0.7cm]  (dNc) edge [out=120, in=60] node {} (dNc);
    \draw[-, min distance=0.5cm]  (dm2) edge [out=120, in=60] node {} (dm2);
    \draw[-] (dm1.east) to (dm2.west);
    \draw[-] (dm2.east) to (dcdot.west);
    \draw[-] (dcdot.east) to (dmp.west);
    \draw[-, min distance=0.7cm]  (dmp) edge [out=120, in=60] node {} (dmp);
    \draw[-] (dmp.east) to (dNc.west);
    \draw[-] ( dm1.north) to [bend left=10] (dNf1.west);
    \draw[-] ( dNf2.west) to [bend left=10] (dm1.south);
    \draw[-] ( dNf1.south) to (dNf2.north);
    
    \node at (-0.4+13,2) {\Large (b)};

    %blue box
    \coordinate (b1) at (12,-0.9) {}; 
    \coordinate (b2) at (12, 0.9) {};
    \coordinate (b3) at (19, 0.9) {};
    \coordinate (b4) at (19,-0.9) {}; 

    \draw[-,blue] (b1) to (b2) to (b3) to (b4) to (b1);

    \node at (15.5,-1.5) {\Large \textcolor{blue}{$\CC_p[U(\WN)]$}};

    % vertical arrows
    \node at (6.3-1,+3.5-1) {\Large $\Downarrow$};
    \node at (6.3-2,+3.5-1) {\Large deconf};

    \node at (15.3,+3.5-1) {\Large $\Uparrow$};
    \node at (15.3+0.8,+3.5-1) {\Large conf};
    
    % electric
    \node[vertex] (kNc) at (5,+5) [shape=circle,draw=black,minimum size=2em] {$N$};
    \node[vertex] (keNf1) at (7,0.8+5) [shape=rectangle,draw=black,minimum height=2em, minimum width=2em] {$F$};
    \node[vertex] (keNf2) at (7,-0.8+5) [shape=rectangle,draw=black,minimum height=2em, minimum width=2em] {$F$};

    \draw[-, min distance=1cm]  (kNc) edge [out=120, in=60] node {} (kNc);
    \draw[-] ([yshift= 2pt] kNc.east) to ([yshift= 0pt] keNf1.west);
    \draw[-] ([yshift= 0pt] keNf2.west) to ([yshift= -2pt] kNc.east);

    \node at (5-0.5,2+5) {\Large (e)};

    % magnetic
    \node[vertex] (kmNc) at (14,+5) [shape=circle,draw=black,minimum size=2em] {$\WN$};
    \node[vertex] (kmNf1) at (16,0.8+5) [shape=rectangle,draw=black,minimum height=2em, minimum width=2em] {$F$};
    \node[vertex] (kmNf2) at (16,-0.8+5) [shape=rectangle,draw=black,minimum height=2em, minimum width=2em] {$F$};

    \draw[-, min distance=1cm]  (kmNc) edge [out=120, in=60] node {} (kmNc);
    \draw[-] ([yshift= 0pt] kmNf1.west) to ([yshift= 2pt] kmNc.east);
    \draw[-] ([yshift= -2pt] kmNc.east) to ([yshift= 0pt] kmNf2.west);
    \draw[-] (kmNf2.north) to (kmNf1.south);

    \node at (14-0.5,2+5) {\Large (m)};

    \node at (10.5,0.3+5) {$KP_\b^+$};
    \node at (10.5,-0.2+5) {\Large $=\!\!=$};

\end{tikzpicture}
\caption{\label{fig: [b]=0} Duality loop for deriving the partition function agreement of $KP_\b^+$ duality for $\b=0$, or $\b=p$. Sequences of $\CA$ and $\CB$ are drawn by solid and dashed arrows respectively. Singlet contributions, FI-parameters, directions and multiplicities of arrows are suppressed for simplicity.}
\end{figure}
\medskip\noindent{\bf $\bullet$ $ \b=0$ :}
Consider quiver (a) in Figure~\ref{fig: [b]=0} for $\b=0$, obtained by deconfining the adjoint in (e). Successive applications of $\CB$ to the $j$-th node, proceeding from $j=p$ to $j=1$, yield
\begin{equation}
    \begin{tikzpicture}[thick,scale=0.7, every node/.style={scale=0.7}]%[shorten >=1pt]
  \tikzset{vertex/.style={circle,fill=white!25,minimum size=2.5em,inner sep=2pt}}
  \tikzset{every loop/.style={}}%removes arrow head from all loops.

% Theory A'
    \node[vertex] (A'n) at (0,0) [shape=circle,draw=black,minimum size=2.5em] {\scriptsize $m_{j\texttt{-}\!1}$};
    \node[vertex] (A'2n) at (1.5,0) [shape=circle,draw=black,minimum size=2.5em] {\scriptsize $m_{j}$};
    \node[vertex] (A'm) at (3,0) [shape=circle,draw=black,minimum size=2.5em] {\scriptsize $\Wm_{p\texttt{-}j}\!$};
    \node[vertex] (A'cd1) at (-1.5,0) {$\cdots$};
    \node[vertex] (A'cd2) at (4.5,0) {$\cdots$};
    \node[vertex] (A'F1) at (6,1) [shape=rectangle,draw=black,minimum height=2.5em, minimum width=2.5em] {$F$};
    \node[vertex] (A'F2) at (6,-1) [shape=rectangle,draw=black,minimum height=2.5em, minimum width=2.5em] {$F$};

    \draw[-to, min distance=1cm]  (A'n) edge [out=120, in=60] node {} (A'n);
    \draw[->-=.5] ([yshift= 2pt] A'cd1.east) to ([yshift= 2pt] A'n.west);
    \draw[->-=.5] ([yshift= -2pt] A'n.west) to ([yshift= -2pt] A'cd1.east);
    \draw[->-=.5] ([yshift= 2pt] A'n.east) to ([yshift= 2pt] A'2n.west);
    \draw[->-=.5] ([yshift= -2pt] A'2n.west) to ([yshift= -2pt] A'n.east);
    \draw[->-=.5] ([yshift= 2pt] A'm.west) to ([yshift= 2pt] A'2n.east);
    \draw[->-=.5] ([yshift= -2pt] A'2n.east) to ([yshift= -2pt] A'm.west);
    \draw[->-=.5] ([yshift= 2pt] A'm.east) to ([yshift= 2pt] A'cd2.west);
    \draw[->-=.5] ([yshift= -2pt] A'cd2.west) to ([yshift= -2pt] A'm.east);
    
    \draw[->-=.5] ([yshift= 0pt] A'F1.west) to [bend right=20] ([xshift= -6pt,yshift=13pt] A'm.east);
    \draw[->-=.5] ([xshift= -6pt,yshift=-13pt] A'm.east) to [bend right=20] ([yshift= 0pt] A'F2.west);
    
    \draw[->-=.5] ([xshift= -5pt,yshift=13pt] A'2n.east) to [bend left=25] ([yshift= 5pt] A'F1.west);
    \draw[->-=.5] ([yshift= -5pt] A'F2.west) to [bend left=25] ([xshift= -5pt,yshift=-13pt] A'2n.east);

    \node at (0,1.3) {\scriptsize $2\texttt{-}2\D$};
    \node at (-0.7,0.4) {\scriptsize $\D$};
    \node at (0.8,0.4) {\scriptsize $\D$};
    \node at (2.3,0.4) {\scriptsize $1\texttt{-}\D$};
    \node at (3.8,0.4) {\scriptsize $\D$}; 
    \node at (6,0) {\scriptsize $\cdots$};

    \node at (3.6,0.9)
    {\begin{tikzpicture}
      \node [rotate=25] {\scriptsize $(j\texttt{+}2)\D\texttt{-}r$};    
    \end{tikzpicture}
    };
    \node at (2.8,1.3)
    {\begin{tikzpicture}
      \node [rotate=15] { $r \texttt{+} (p\texttt{-}j)\D$};    
    \end{tikzpicture}
    };

    \node at (0,-2+0.3) {\scriptsize $\t_{j\texttt{-}1}$};
    
    \node at (1.5,-2+0.3) {\scriptsize $\k_j$};
    
    \node at (3,-2+0.3) {\scriptsize $\texttt{-}\t_{j\texttt{+}1,p}$};
    \node at (3,-2) {\tiny $\texttt{+}(p\texttt{-}j\texttt{-}1)\D\texttt{+}1$};

    \node at (8,0.5) {$\CB$};
    \node at (8,0) {\LARGE $=\!\!=$};

    % Theory B'
    \node[vertex] (B'n) at (11,0) [shape=circle,draw=black,minimum size=2.5em]{\scriptsize $m_{j\texttt{-}\!1}$};
    \node[vertex] (B'2m) at (12.5,0) [shape=circle,draw=black,minimum size=2.5em] {\scriptsize $\Wm_{p\texttt{-}\!j\texttt{+}\!1\!}$};
    \node[vertex] (B'm) at (14,0) [shape=circle,draw=black,minimum size=2.5em] {\scriptsize $\Wm_{p\texttt{-}j}$};
    \node[vertex] (B'cd1) at (9.5,0) {$\cdots$};
    \node[vertex] (B'cd2) at (15.5,0) {$\cdots$};
    \node[vertex] (B'F1) at (17,1) [shape=rectangle,draw=black,minimum height=2.5em, minimum width=2.5em] {$F$};
    \node[vertex] (B'F2) at (17,-1) [shape=rectangle,draw=black,minimum height=2.5em, minimum width=2.5em] {$F$};

    \draw[->-=.5] ([yshift= 2pt] B'cd1.east) to ([yshift= 2pt] B'n.west);
    \draw[->-=.5] ([yshift= -2pt] B'n.west) to ([yshift= -2pt] B'cd1.east);
    \draw[->-=.5] ([yshift= 2pt] B'2m.west) to ([yshift= 2pt] B'n.east);
    \draw[->-=.5] ([yshift= -2pt] B'n.east) to ([yshift= -2pt] B'2m.west);
    \draw[->-=.5] ([yshift= 2pt] B'2m.east) to ([yshift= 2pt] B'm.west);
    \draw[->-=.5] ([yshift= -2pt] B'm.west) to ([yshift= -2pt] B'2m.east);
    \draw[-to, min distance=0.8cm]  (B'm) edge [out=120, in=60] node {} (B'm);
    \draw[->-=.5] ([yshift= 2pt] B'm.east) to ([yshift= 2pt] B'cd2.west);
    \draw[->-=.5] ([yshift= -2pt] B'cd2.west) to ([yshift= -2pt] B'm.east);
    \draw[->-=.5] ([yshift= 5pt] B'F1.west) to [bend right=25] ([xshift= -5pt,yshift=13pt] B'2m.east);
    \draw[->-=.5] ([xshift= -5pt,yshift=-13pt] B'2m.east) to [bend right=25] ([yshift= -5pt] B'F2.west);
    \draw[->-=.5] ([xshift= 3pt,yshift=0pt] B'n.north) to [bend left=20] ([yshift= 10pt] B'F1.west);
    \draw[->-=.5] ([yshift= -10pt] B'F2.west) to [bend left=20] ([xshift= 3pt,yshift=0pt] B'n.south);
    
    \draw[->-=.6] ([xshift= 7pt] B'F2.north) to ([xshift= 7pt] B'F1.south);

    \node at (-0.7+11,0.4) {\scriptsize $\D$};
    \node at (0.8+11,0.4) {\scriptsize $1\texttt{-}\D$};
    \node at (2.3+11,0.4) {\scriptsize $\D$};
    \node at (3.8+11,0.4) {\scriptsize $\D$};
    \node at (3.8+11,0.9) {\scriptsize $2\texttt{-}2\D$};
    \node at (18.3,0) {\scriptsize $2r \texttt{+} 2(p\texttt{-}j)\D$};
    \node at (17-0.1,0) {\scriptsize $\cdots$}; 
    
    \node at (17,-2) {$+\big\{ 2-2\k_j \big\}$};
    
    \node at (4.15+9,0.9)
    {\begin{tikzpicture}
      \node [rotate=29] { $(j\texttt{+}1)\D\texttt{-}r$};    
    \end{tikzpicture}
    };
    \node at (2.8+9.3,1.3)
    {\begin{tikzpicture}
      \node [rotate=15] { $r \texttt{+} (p\texttt{-}j\texttt{+}1)\D$};    
    \end{tikzpicture}
    };

    \node at (0+11,-2) {\scriptsize $\k_{j\texttt{-}1}$};
    
    \node at (1.5+11,-2) {\scriptsize $\texttt{-}\t_{j,p}$};
    \node at (1.5+11,-2-0.3) {\tiny $\texttt{+}(p\texttt{-}j)\D\texttt{+}1$};

    \node at (3+11,-2) {\scriptsize $\t_{j}$};

\end{tikzpicture}
\end{equation}
where $\Wm_j=jF+m_{p-j}-N-j$ and $\k_j=\t_{j,p}-(p-j)\D$, with $\t_{j,p}\equiv\sum_{i=j}^p\t_i$. Importantly, the initial R-charge $2$ condition \eqref{eq: FI cond},
\begin{align}
F(1-r)-2\D(N-1)+\t=2
\,,
\end{align}
automatically ensures the monopole R-charge 2 condition at each subsequent $\CB$ step:
\begin{align}
m_{j-1}(1-\D)+\Wm_{p-j}\D+F(1-r-(p-j)\D)-m_j+1+\k_j=2
\,,
\end{align}
where $m_0=\Wm_0\equiv0$. The resulting linear tail, with its last node of rank $\Wm_p=pF-N-p=\WN$, forms $\CC_p[U(\WN)]$. Its confinement into an R-charge $2\D$ adjoint then yields the expected quiver (m). Note that $\tilde\t_p=-\t+2p\D-[\WN-1]_p\D$, which gives the desired FI-parameter $\tilde{\t}=-\t+2p\D$ in (m). The remaining singlet contributions precisely match those of the singlets $V_i^-$ coupled to the dual monopoles in the superpotential:
\begin{align}
    \big\{ 2-2\k_j \big\}_{j=1}^{p}
    =
    \big\{ F(1-r) - 2\D(N-1 - j) - \t \big\}_{j=0}^{p-1}
    \,.
\end{align}

\medskip\noindent{\bf $\bullet$ $ \b=p$ :}
Similarly, when $\b=p$, applying $\CA$ from $j=p$ to $j=1$ instead is given by
\begin{equation}
    \begin{tikzpicture}[thick,scale=0.7, every node/.style={scale=0.7}]%[shorten >=1pt]
  \tikzset{vertex/.style={circle,fill=white!25,minimum size=2.5em,inner sep=2pt}}
  \tikzset{every loop/.style={}}%removes arrow head from all loops.

% Theory A'
    \node[vertex] (A'n) at (0,0) [shape=circle,draw=black,minimum size=2.5em] {\scriptsize $m_{j\texttt{-}\!1}$};
    \node[vertex] (A'2n) at (1.5,0) [shape=circle,draw=black,minimum size=2.5em] {\scriptsize $m_{j}$};
    \node[vertex] (A'm) at (3,0) [shape=circle,draw=black,minimum size=2.5em] {\scriptsize $\Wm_{p\texttt{-}j}\!$};
    \node[vertex] (A'cd1) at (-1.5,0) {$\cdots$};
    \node[vertex] (A'cd2) at (4.5,0) {$\cdots$};
    \node[vertex] (A'F1) at (6,1) [shape=rectangle,draw=black,minimum height=2.5em, minimum width=2.5em] {$F$};
    \node[vertex] (A'F2) at (6,-1) [shape=rectangle,draw=black,minimum height=2.5em, minimum width=2.5em] {$F$};

    \draw[-to, min distance=1cm]  (A'n) edge [out=120, in=60] node {} (A'n);
    \draw[->-=.5] ([yshift= 2pt] A'cd1.east) to ([yshift= 2pt] A'n.west);
    \draw[->-=.5] ([yshift= -2pt] A'n.west) to ([yshift= -2pt] A'cd1.east);
    \draw[->-=.5] ([yshift= 2pt] A'n.east) to ([yshift= 2pt] A'2n.west);
    \draw[->-=.5] ([yshift= -2pt] A'2n.west) to ([yshift= -2pt] A'n.east);
    \draw[->-=.5] ([yshift= 2pt] A'm.west) to ([yshift= 2pt] A'2n.east);
    \draw[->-=.5] ([yshift= -2pt] A'2n.east) to ([yshift= -2pt] A'm.west);
    \draw[->-=.5] ([yshift= 2pt] A'm.east) to ([yshift= 2pt] A'cd2.west);
    \draw[->-=.5] ([yshift= -2pt] A'cd2.west) to ([yshift= -2pt] A'm.east);
    
    \draw[->-=.5] ([yshift= 0pt] A'F1.west) to [bend right=20] ([xshift= -6pt,yshift=13pt] A'm.east);
    \draw[->-=.5] ([xshift= -6pt,yshift=-13pt] A'm.east) to [bend right=20] ([yshift= 0pt] A'F2.west);
    
    \draw[->-=.5] ([xshift= -5pt,yshift=13pt] A'2n.east) to [bend left=25] ([yshift= 5pt] A'F1.west);
    \draw[->-=.5] ([yshift= -5pt] A'F2.west) to [bend left=25] ([xshift= -5pt,yshift=-13pt] A'2n.east);

    \node at (0,1.3) {\scriptsize $2\texttt{-}2\D$};
    \node at (-0.7,0.4) {\scriptsize $\D$};
    \node at (0.8,0.4) {\scriptsize $\D$};
    \node at (2.3,0.4) {\scriptsize $1\texttt{-}\D$};
    \node at (3.8,0.4) {\scriptsize $\D$}; 
    \node at (6,0) {\scriptsize $\cdots$};

    \node at (3.6,0.9)
    {\begin{tikzpicture}
      \node [rotate=25] {\scriptsize $(j\texttt{+}2)\D\texttt{-}r$};    
    \end{tikzpicture}
    };
    \node at (2.8,1.3)
    {\begin{tikzpicture}
      \node [rotate=15] { $r \texttt{+} (p\texttt{-}j)\D$};    
    \end{tikzpicture}
    };

    \node at (0,-2+0.3) {\scriptsize $\t_{j\texttt{-}1}$};
    
    \node at (1.5,-2+0.3) {\scriptsize $\t_{j,p}$};
    
    \node at (3,-2+0.3) {\scriptsize $\texttt{-}\t_{j\texttt{+}1,p}$};

    \node at (8,0.5) {$\CA$};
    \node at (8,0) {\LARGE $=\!\!=$};

    % Theory B'
    \node[vertex] (B'n) at (11,0) [shape=circle,draw=black,minimum size=2.5em]{\scriptsize $m_{j\texttt{-}\!1}$};
    \node[vertex] (B'2m) at (12.5,0) [shape=circle,draw=black,minimum size=2.5em] {\scriptsize $\Wm_{p\texttt{-}\!j\texttt{+}\!1\!}$};
    \node[vertex] (B'm) at (14,0) [shape=circle,draw=black,minimum size=2.5em] {\scriptsize $\Wm_{p\texttt{-}j}$};
    \node[vertex] (B'cd1) at (9.5,0) {$\cdots$};
    \node[vertex] (B'cd2) at (15.5,0) {$\cdots$};
    \node[vertex] (B'F1) at (17,1) [shape=rectangle,draw=black,minimum height=2.5em, minimum width=2.5em] {$F$};
    \node[vertex] (B'F2) at (17,-1) [shape=rectangle,draw=black,minimum height=2.5em, minimum width=2.5em] {$F$};

    \draw[->-=.5] ([yshift= 2pt] B'cd1.east) to ([yshift= 2pt] B'n.west);
    \draw[->-=.5] ([yshift= -2pt] B'n.west) to ([yshift= -2pt] B'cd1.east);
    \draw[->-=.5] ([yshift= 2pt] B'2m.west) to ([yshift= 2pt] B'n.east);
    \draw[->-=.5] ([yshift= -2pt] B'n.east) to ([yshift= -2pt] B'2m.west);
    \draw[->-=.5] ([yshift= 2pt] B'2m.east) to ([yshift= 2pt] B'm.west);
    \draw[->-=.5] ([yshift= -2pt] B'm.west) to ([yshift= -2pt] B'2m.east);
    \draw[-to, min distance=0.8cm]  (B'm) edge [out=120, in=60] node {} (B'm);
    \draw[->-=.5] ([yshift= 2pt] B'm.east) to ([yshift= 2pt] B'cd2.west);
    \draw[->-=.5] ([yshift= -2pt] B'cd2.west) to ([yshift= -2pt] B'm.east);
    \draw[->-=.5] ([yshift= 5pt] B'F1.west) to [bend right=25] ([xshift= -5pt,yshift=13pt] B'2m.east);
    \draw[->-=.5] ([xshift= -5pt,yshift=-13pt] B'2m.east) to [bend right=25] ([yshift= -5pt] B'F2.west);
    \draw[->-=.5] ([xshift= 3pt,yshift=0pt] B'n.north) to [bend left=20] ([yshift= 10pt] B'F1.west);
    \draw[->-=.5] ([yshift= -10pt] B'F2.west) to [bend left=20] ([xshift= 3pt,yshift=0pt] B'n.south);
    
    \draw[->-=.6] ([xshift= 7pt] B'F2.north) to ([xshift= 7pt] B'F1.south);

    \node at (-0.7+11,0.4) {\scriptsize $\D$};
    \node at (0.8+11,0.4) {\scriptsize $1\texttt{-}\D$};
    \node at (2.3+11,0.4) {\scriptsize $\D$};
    \node at (3.8+11,0.4) {\scriptsize $\D$};
    \node at (3.8+11,0.9) {\scriptsize $2\texttt{-}2\D$};
    \node at (18.3,0) {\scriptsize $2r \texttt{+} 2(p\texttt{-}j)\D$};
    \node at (17-0.1,0) {\scriptsize $\cdots$}; 
    
    \node at (17,-2) {$+\big\{ R_j \pm \t_{j,p} \big\}$};
    
    \node at (4.15+9,0.9)
    {\begin{tikzpicture}
      \node [rotate=29] { $(j\texttt{+}1)\D\texttt{-}r$};    
    \end{tikzpicture}
    };
    \node at (2.8+9.3,1.3)
    {\begin{tikzpicture}
      \node [rotate=15] { $r \texttt{+} (p\texttt{-}j\texttt{+}1)\D$};    
    \end{tikzpicture}
    };

    \node at (0+11,-2) {\scriptsize $\t_{j\texttt{-}1,p}$};
    
    \node at (1.5+11,-2) {\scriptsize $\texttt{-}\t_{j,p}$};

    \node at (3+11,-2) {\scriptsize $\t_{j}$};

\end{tikzpicture}
\end{equation}
with $\Wm_j = j F + m_{p-j} - N$ and $R_j = F(1-r)-N+1 + (N + m_{j-1}-m_j)(1-\D)$. After the dualization, the last node has rank $\Wm_p=pF-N=\WN$, forming a $\CC_p[U(\WN)]$ tail whose confinement yields an adjoint of R-charge $2\D$ and the expected quiver (m). The FI-parameter $\tilde{\t}_p=-\t_{1,p}=-\t-[\WN-1]_p\D$ also reduces to the expected $\tilde{\t}=-\t$ upon confinement. Given that
\begin{align}
    F(1-r) - 2\D (N-1 - p) + \t = 2
    \,,
\end{align}
$p$ of the $2p$ singlet contributions cancel, leaving precisely the expected result
\begin{align}
    \big\{
    R_j \pm \t_{j,p}
    \big\}_{j=1}^{p}
    =
    \big\{
    F(1-r) - 2\D(N-1-j) - \t
    \big\}_{j=0}^{p-1}
    \,.
\end{align}

\subsubsection{For \texorpdfstring{$0<\b<p$ }{0<b<p} }
\begin{figure}[thbp]
\centering
\begin{tikzpicture}[thick,scale=0.65, every node/.style={scale=0.6}]%[shorten >=1pt]%[shorten >=1pt]
  \tikzset{vertex/.style={circle,fill=white!25,minimum size=12pt,inner sep=2pt}}
  \tikzset{every loop/.style={}}%removes arrow head from all loops.
    
    % LHS Theory
    \node[vertex] (m1) at (0,0) [shape=circle,draw=black,minimum size=2em] {$m_1$};
    \node[vertex] (m2) at (1.5,0) [shape=circle,draw=black,minimum size=2em] {$m_2$};
    \node (ecdot) at (3,0) {$\cdots$};
    \node[vertex] (mp-1) at (4.5,0) [shape=circle,draw=black,minimum size=2em] {\tiny$m_{p\texttt{-}1}$};
    \node[vertex] (Nc) at (6,0) [shape=circle,draw=black,minimum size=2em] {$N$};
    \node[vertex] (eNf1) at (8,0.8) [shape=rectangle,draw=black,minimum height=2em, minimum width=2em] {$F$};
    \node[vertex] (eNf2) at (8,-0.8) [shape=rectangle,draw=black,minimum height=2em, minimum width=2em] {$F$};

    \draw[-to, min distance=1cm]  (m1) edge [out=120, in=60] node {} (m1);
    \draw[-to, min distance=1cm]  (m2) edge [out=120, in=60] node {} (m2);
    \draw[->-=.5] ([yshift= 2pt] m1.east) to ([yshift= 2pt] m2.west);
    \draw[->-=.5] ([yshift= -2pt] m2.west) to ([yshift= -2pt] m1.east);
    \draw[->-=.5] ([yshift= 2pt] m2.east) to ([yshift= 2pt] ecdot.west);
    \draw[->-=.5] ([yshift= -2pt] ecdot.west) to ([yshift= -2pt] m2.east);
    \draw[->-=.5] ([yshift= 2pt] ecdot.east) to ([yshift= 2pt] mp-1.west);
    \draw[->-=.5] ([yshift= -2pt] mp-1.west) to ([yshift= -2pt] ecdot.east);
    \draw[-to, min distance=1cm]  (mp-1) edge [out=120, in=60] node {} (mp-1);
    \draw[->-=.5] ([yshift= 2pt] mp-1.east) to ([yshift= 2pt] Nc.west);
    \draw[->-=.5] ([yshift= -2pt] Nc.west) to ([yshift= -2pt] mp-1.east);
    \draw[->-=.5] ([yshift= 2pt] Nc.east) to ([yshift= 0pt] eNf1.west);
    \draw[->-=.5] ([yshift= 0pt] eNf2.west) to ([yshift= -2pt] Nc.east);

    \node at (-0.5,2) {(a)};
    
    \node at (0,1.3) {$2\texttt{-}2\D$};
    \node at (1.5,1.3) {$2\texttt{-}2\D$};
    \node at (4.5,1.3) {$2\texttt{-}2\D$};
    \node at (0.8,0.4) {$\D$};
    \node at (2.3,0.4) {$\D$};
    \node at (3.7,0.4) {$\D$};
    \node at (5.3,0.4) {$\D$};
    \node at (6.8,0.8) {$r$};

    \node at (0,-0.8) {$\t_1$};
    \node at (1.5,-0.8) {$\t_2$};
    \node at (4.5,-0.8) {$\t_{p\texttt{-}1}$};
    \node at (6,-0.8) {$\t_p$};

    \node at (7.5,1.6) {$+\{2\D j\}_{j=1}^{\min(\a,p-\a)}$};

    \node at (11-0.3,1.1) {\scriptsize Sequential dual};
    \node at (11-0.3,0.7) {\scriptsize by $\CA$ and $\CB$};
    \node at (11-0.3,0.3) {\scriptsize then confine};
    \node at (11-0.3,-0.2) {\Large $=\!\!=$};
    
    % RHS Theory
    \node[vertex] (dmp-1) at (16.5-0.1-3,0) [shape=circle,draw=black,minimum size=2em] {$M$};
    \node[vertex] (dNc) at (18-0.1-3,0) [shape=circle,draw=black,minimum size=2em] {$\WN$};
    \node[vertex] (dNf1) at (20-0.1-3,0.8) [shape=rectangle,draw=black,minimum height=2em, minimum width=2em] {$F$};
    \node[vertex] (dNf2) at (20-0.1-3,-0.8) [shape=rectangle,draw=black,minimum height=2em, minimum width=2em] {$F$};

    \draw[-to, min distance=1cm]  (dmp-1) edge [out=120, in=60] node {} (dmp-1);
    \draw[-to, min distance=1cm]  (dNc) edge [out=120, in=60] node {} (dNc);
    \draw[->-=.5] ([yshift= 2pt] dNc.west) to ([yshift= 2pt] dmp-1.east);
    \draw[->-=.5] ([yshift= -2pt] dmp-1.east) to ([yshift= -2pt] dNc.west);
    \draw[->-=.5] ([yshift= 0pt] dNf1.west) to ([yshift= 2pt] dNc.east);
    \draw[->-=.5] ([yshift= -2pt] dNc.east) to ([yshift= -2pt] dNf2.west);
    \draw[->-=.7] ([xshift= -5pt] dNf2.north) to ([xshift= -5pt] dNf1.south);
    \draw[->-=.7] ([xshift= 5pt] dNf2.north) to ([xshift= 5pt] dNf1.south);
    
    \node at (-0.4+13,2) {(b)};

     \node at (16.5-0.1-3,1.3) {$2\D$};
     \node at (18-0.1-3,1.3) {$2\D$};
    \node at (17.3-0.1-3,0.4) {$2\D$};
    \node at (19-0.1-3,0.8) {\scriptsize $2\D\texttt{-}r$};
    \node at (20-0.1-0.1-3,-0.1) {\tiny$\cdot$};
    \node at (20-0.1-3,-0.1) {\tiny$\cdot$};
    \node at (20+0.1-0.1-3,-0.1) {\tiny$\cdot$};
    \node at (21.3-0.1-3,0) {$2r \texttt{+} 2j\D$};
    \node at (21.5-0.1-3,-0.4) {\scriptsize $j=0,\cdots,p\texttt{-}1$};

    \node at (20.2-0.1-3,2.2) {$+\{2\D (M+j)\}_{j=1}^p$};
    \node at (20.2-0.1-3,1.6) {$+\{R_0 + 2\D j - \t\}_{j=0}^{p\texttt{-}1}$};

    \node at (16.5-0.1-3,-0.8) {\tiny $\WN(1\texttt{-}2\D)$};
    \node at (18-0.1-3,-0.8) {\tiny $\tilde{\t}\texttt{-}M(1\texttt{-}2\D)$};

    % vertical arrows
    \node at (6.3-1,+3.5) 
    {\begin{tikzpicture}
      \node [rotate=90] {\LARGE\textcolor{black}{$=\!\!=$}};    
    \end{tikzpicture}
    };
    \node at (6.3-2.3,+3.5) {(de)confine};

    \node at (15.3,+3.5) 
    {\begin{tikzpicture}
      \node [rotate=-90] {\LARGE\textcolor{black}{$=\!\!=$}};    
    \end{tikzpicture}
    };
    \node at (15.3+1.3,+3.5) {rank-reducing};
    
    % electric
    \node[vertex] (kNc) at (5,+6) [shape=circle,draw=black,minimum size=2em] {$N$};
    \node[vertex] (keNf1) at (7,0.8+6) [shape=rectangle,draw=black,minimum height=2em, minimum width=2em] {$F$};
    \node[vertex] (keNf2) at (7,-0.8+6) [shape=rectangle,draw=black,minimum height=2em, minimum width=2em] {$F$};

    \draw[-to, min distance=1cm]  (kNc) edge [out=120, in=60] node {} (kNc);
    \draw[->-=.5] ([yshift= 2pt] kNc.east) to ([yshift= 0pt] keNf1.west);
    \draw[->-=.5] ([yshift= 0pt] keNf2.west) to ([yshift= -2pt] kNc.east);

    \node at (5-0.5,2+6) {(e)};

    \node at (5,1.3+6) {$2\D$};
    \node at (5,-0.8+6) {$\t$};
    \node at (6,1+6) {$r$};

    % magnetic
    \node[vertex] (kmNc) at (14,+6) [shape=circle,draw=black,minimum size=2em] {$\WN$};
    \node[vertex] (kmNf1) at (16,0.8+6) [shape=rectangle,draw=black,minimum height=2em, minimum width=2em] {$F$};
    \node[vertex] (kmNf2) at (16,-0.8+6) [shape=rectangle,draw=black,minimum height=2em, minimum width=2em] {$F$};

    \draw[-to, min distance=1cm]  (kmNc) edge [out=120, in=60] node {} (kmNc);
    \draw[->-=.5] ([yshift= 0pt] kmNf1.west) to ([yshift= 2pt] kmNc.east);
    \draw[->-=.5] ([yshift= -2pt] kmNc.east) to ([yshift= 0pt] kmNf2.west);
    \draw[->-=.7] ([xshift= -5pt] kmNf2.north) to ([xshift= -5pt] kmNf1.south);
    \draw[->-=.7] ([xshift= 5pt] kmNf2.north) to ([xshift= 5pt] kmNf1.south);

    \node at (14-0.5,2+6) {(m)};
    
    \node at (16-0.1,-0.1+6) {\tiny$\cdot$};
    \node at (16,-0.1+6) {\tiny$\cdot$};
    \node at (16+0.1,-0.1+6) {\tiny$\cdot$};
    \node at (14,1.3+6) {$2\D$};
    \node at (15,1+6) {\scriptsize $2\D\texttt{-}r$};
    \node at (14,-0.8+6) {$\tilde{\t}$};

    \node at (17.3,0+6) {$2r \texttt{+} 2j\D$};
    \node at (17.5,-0.4+6) {\scriptsize $j=0,\cdots,p\texttt{-}1$};

    \node at (18.2,1.5+6) {$+\{R_0 + 2\D j - \t\}_{j=0}^{p\texttt{-}1}$};

    \node at (10.5,0.3+6) {$KP_\b^+$};
    \node at (10.5,-0.7+6) {\eqref{eq: KPb quiver}};
    \node at (10.5,-0.2+6) {\Large $=\!\!=$};
    
\end{tikzpicture}
\caption{\label{fig: <b<} Duality loop for deriving the partition function agreement of $KP_\b^+$ for $0<\b<p$ cases.}
\end{figure}
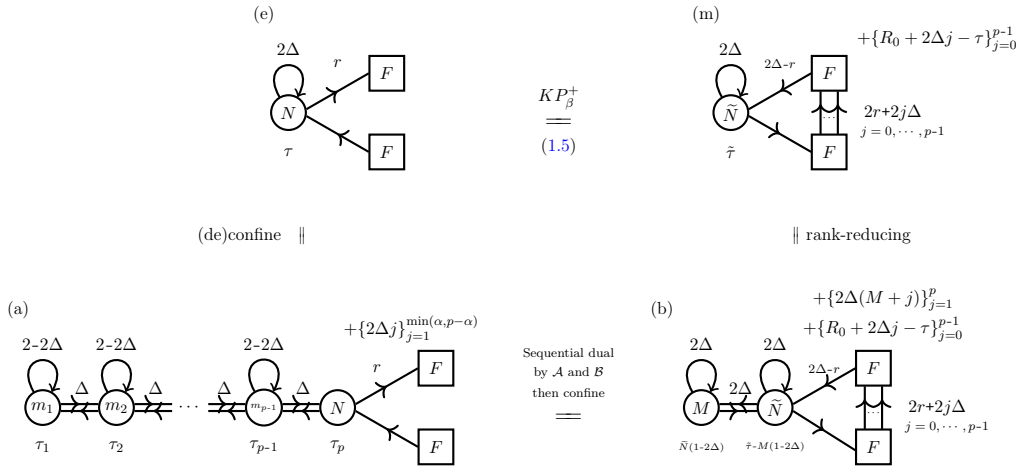

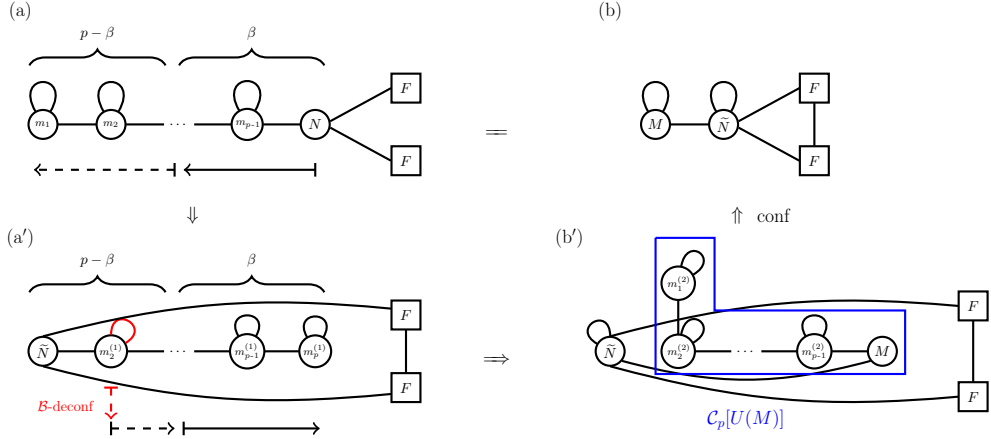
\begin{figure}[thbp]
\centering
\begin{tikzpicture}[thick,scale=0.6, every node/.style={scale=0.5}]%[shorten >=1pt]%[shorten >=1pt]
  \tikzset{vertex/.style={circle,fill=white!25,minimum size=12pt,inner sep=2pt}}
  \tikzset{every loop/.style={}}%removes arrow head from all loops.
    
    % a Theory
    \node at (-0.5,2.5) {\Large (a)};
    
    \node[vertex] (m1) at (0,0) [shape=circle,draw=black,minimum size=2em] {\scriptsize$m_1$};
    \node[vertex] (m2) at (1.5,0) [shape=circle,draw=black,minimum size=2em] {\scriptsize$m_2$};
    \node (ecdot) at (3,0) {$\cdots$};
    \node[vertex] (mp-1) at (4.5,0) [shape=circle,draw=black,minimum size=2em] {\scriptsize$m_{p\texttt{-}1}\!$};
    \node[vertex] (Nc) at (6,0) [shape=circle,draw=black,minimum size=2em] {$N$};
    \node[vertex] (eNf1) at (8,0.8) [shape=rectangle,draw=black,minimum height=2em, minimum width=2em] {$F$};
    \node[vertex] (eNf2) at (8,-0.8) [shape=rectangle,draw=black,minimum height=2em, minimum width=2em] {$F$};

    \draw[-, min distance=1cm]  (m1) edge [out=120, in=60] node {} (m1);
    \draw[-, min distance=1cm]  (m2) edge [out=120, in=60] node {} (m2);
    \draw[-] (m1.east) to (m2.west);
    \draw[-] (m2.east) to (ecdot.west);
    \draw[-] (ecdot.east) to (mp-1.west);
    \draw[-, min distance=1cm]  (mp-1) edge [out=120, in=60] node {} (mp-1);
    \draw[-] (mp-1.east) to (Nc.west);
    \draw[-] ([yshift= 2pt] Nc.east) to ([yshift= 0pt] eNf1.west);
    \draw[-] ([yshift= 0pt] eNf2.west) to ([yshift= -2pt] Nc.east);

    \draw[decorate, decoration={brace, amplitude=6pt}] (-0.3,1.3) -- (2.7,1.3);
    \draw[decorate, decoration={brace, amplitude=6pt}] (3,1.3) -- (6.2,1.3);
    \node at (1.2,2) {$p-\b$};
    \node at (4.6,2) {$\b$};

    \node at (10,-0.2-5) {\Large $=\!\!\Rightarrow$};

    %Arrows
    \coordinate (an) at (6,-1+0.15);
    \coordinate (as) at (6,-1-0.15);
    \coordinate (ae) at (6,-1);
    \coordinate (aw) at (3+0.1,-1);

    \coordinate (an1) at (3-0.1,-1+0.15);
    \coordinate (as1) at (3-0.1,-1-0.15);
    \coordinate (ae1) at (3-0.1,-1);
    \coordinate (aw1) at (-0.2,-1);

    \draw[-] (an) to (as);
    \draw[-][-to] (ae) to (aw);

    \draw[-] (an1) to (as1);
    \draw[dashed][-to] (ae1) to (aw1);

    % b Theory
    \node at (-0.5+13,2.5) {\Large (b)};
    
    \node[vertex] (bN) at (4.5+9,0) [shape=circle,draw=black,minimum size=2em] {$M$};
    \node[vertex] (bM) at (6+9,0) [shape=circle,draw=black,minimum size=2em] {$\WN$};
    \node[vertex] (bNf1) at (8+9,0.8) [shape=rectangle,draw=black,minimum height=2em, minimum width=2em] {$F$};
    \node[vertex] (bNf2) at (8+9,-0.8) [shape=rectangle,draw=black,minimum height=2em, minimum width=2em] {$F$};

    \draw[-, min distance=1cm]  (bN) edge [out=120, in=60] node {} (bN);
    \draw[-, min distance=1cm]  (bM) edge [out=120, in=60] node {} (bM);
    \draw[-] (bN.east) to (bM.west);
    \draw[-] ([yshift= 2pt] bM.east) to ([yshift= 0pt] bNf1.west);
    \draw[-] ([yshift= 0pt] bNf2.west) to ([yshift= -2pt] bM.east);
    \draw[-] (bNf1.south) to (bNf2.north);

    % a' Theory
    \node at (-0.5,2.5-5) {\Large (a$'$)};
    
    \node[vertex] (bm1) at (0,-5) [shape=circle,draw=black,minimum size=2em] {$\WN$};
    \node[vertex] (bm2) at (1.5,0-5) [shape=circle,draw=black,minimum size=2em] {\scriptsize $\!m_2^{(1)}\!$};
    \node (becdot) at (3,0-5) {$\cdots$};
    \node[vertex] (bmp-1) at (4.5,0-5) [shape=circle,draw=black,minimum size=2em] {\scriptsize$\!m_{p\texttt{-}1}^{(1)}\!$};
    \node[vertex] (bNc) at (6,0-5) [shape=circle,draw=black,minimum size=2em] {\scriptsize$\!m_p^{(1)}\!$};
    \node[vertex] (beNf1) at (8,0.8-5) [shape=rectangle,draw=black,minimum height=2em, minimum width=2em] {$F$};
    \node[vertex] (beNf2) at (8,-0.8-5) [shape=rectangle,draw=black,minimum height=2em, minimum width=2em] {$F$};

    \draw[-, min distance=0.7cm,red,thick]  (bm2) edge [out=90, in=30] node {} (bm2);
    \draw[-, min distance=0.7cm]  (bNc) edge [out=120, in=60] node {} (bNc);

    \draw[-] (bm1.east) to (bm2.west);
    \draw[-] (bm2.east) to (becdot.west);
    \draw[-] (becdot.east) to (bmp-1.west);
    \draw[-, min distance=0.7cm]  (bmp-1) edge [out=120, in=60] node {} (bmp-1);
    \draw[-] (bmp-1.east) to (bNc.west);
    \draw[-] [bend left=10]([yshift= 0pt] bm1.north) to ([yshift= 4pt] beNf1.west);
    \draw[-] [bend left=10]([yshift= -4pt] beNf2.west) to ([yshift= 0pt] bm1.south);
    \draw[-] (beNf1.south) to (beNf2.north);

    \draw[decorate, decoration={brace, amplitude=6pt}] (-0.3,1.3-5) -- (2.7,1.3-5);
    \draw[decorate, decoration={brace, amplitude=6pt}] (3,1.3-5) -- (6.2,1.3-5);
    \node at (1.2,2-5) {$p-\b$};
    \node at (4.6,2-5) {$\b$};

    \node at (10,-0.2) {\Large $=\!\!=$};

    %Arrows
    \coordinate (bw) at (1.5-0.15,-0.8-5);
    \coordinate (be) at (1.5+0.15,-0.8-5);
    \coordinate (bn) at (1.5,-0.8-5);
    \coordinate (bs) at (1.5,-1.5-5);
    \draw[-,red] (bw) to (be);
    \draw[dashed,red][-to] (bn) to (bs);
    \node at (0.5,-6.2) {\textcolor{red}{$\CB$-deconf}};

    \coordinate (bbn) at (1.5,-1.7+0.15-5);
    \coordinate (bbs) at (1.5,-1.7-0.15-5);
    \coordinate (bbw) at (1.5,-1.7-5);
    \coordinate (bbe) at (3-0.1,-1.7-5);
    \draw[-] (bbn) to (bbs);
    \draw[dashed][-to] (bbw) to (bbe);

    \coordinate (bbbn) at (3.1,-1.7+0.15-5);
    \coordinate (bbbs) at (3.1,-1.7-0.15-5);
    \coordinate (bbbw) at (3.1,-1.7-5);
    \coordinate (bbbe) at (6.1,-1.7-5);
    \draw[-] (bbbn) to (bbbs);
    \draw[-to] (bbbw) to (bbbe);

    % b' Theory
    \node at (-0.4+12,2.5-5) {\Large (b$'$)};
    
    \node[vertex] (dm1) at (12.5,0-5) [shape=circle,draw=black,minimum size=2em] {$\WN$};
    \node[vertex] (dm2) at (1.5+12.5,0-5) [shape=circle,draw=black,minimum size=2em] {\scriptsize $m_2^{(2)}\!$};
    \node[vertex] (dm0) at (1.5+12.5,1.5-5) [shape=circle,draw=black,minimum size=2em] {\scriptsize $m_1^{(2)}\!$};
    \node (dcdot) at (3+12.5,0-5) {$\cdots$};
    \node[vertex] (dmp) at (4.5+12.5,0-5) [shape=circle,draw=black,minimum size=2em] {\scriptsize $\!m_{p\texttt{-}1}^{(2)}\!$};
    \node[vertex] (dNc) at (6+12.5,0-5) [shape=circle,draw=black,minimum size=2em] {$M$};
    \node[vertex] (dNf1) at (8+12.5,1-5) [shape=rectangle,draw=black,minimum height=2em, minimum width=2em] {$F$};
    \node[vertex] (dNf2) at (8+12.5,-1-5) [shape=rectangle,draw=black,minimum height=2em, minimum width=2em] {$F$};

    \draw[-, min distance=0.7cm]  (dm1) edge [out=150, in=90] node {} (dm1);    \draw[-, min distance=0.7cm]  (dm0) edge [out=80, in=30] node {} (dm0);
    \draw[-, min distance=0.7cm]  (dm2) edge [out=80, in=30] node {} (dm2);
    \draw[-] ([xshift= -2pt,yshift= -6pt] dm1.east) to [bend right=15] ([xshift= 2pt,yshift= -6pt] dNc.west);
    
    \draw[-] (dm2.east) to (dcdot.west);
    \draw[-] (dm0.south) to (dm2.north);
    \draw[-] (dcdot.east) to (dmp.west);
    \draw[-, min distance=0.7cm]  (dmp) edge [out=120, in=60] node {} (dmp);
    \draw[-] (dmp.east) to (dNc.west);
    \draw[-] ( dm1.north) to [bend left=10] (dNf1.west);
    \draw[-] ( dNf2.west) to [bend left=10] (dm1.south);
    \draw[-] ( dNf1.south) to (dNf2.north);

    %blue box
    \coordinate (b1) at (13.5,2.5-5) {}; 
    \coordinate (b2) at (15-0.2,2.5-5) {}; 
    \coordinate (b3) at (15-0.2,0.9-5) {}; 
    \coordinate (b4) at (19,0.9-5) {}; 
    \coordinate (b5) at (19,-0.5-5) {}; 
    \coordinate (b6) at (13.5,-0.5-5) {}; 
    
    \draw[-,blue] (b1) to (b2) to (b3) to (b4) to (b5) to (b6) to (b1);

    \node at (15.5,-1.5-5) {\Large \textcolor{blue}{$\CC_p[U(M)]$}};

    % vertical arrows
    \node at (4.3-1,-2) {\Large $\Downarrow$};

    \node at (15.3,-2) {\Large $\Uparrow$};
    \node at (15.3+0.8,-2) {\Large conf};
    
\end{tikzpicture}
\caption{\label{fig: a to b} Duality loop from (a) to (b) in Figure \ref{fig: <b<} for $0< \b <p$ cases. Solid and dashed arrows stand for dualizations by $\CA$ and $\CB$ respectively.}
\end{figure}

We now consider the cases $0<\b<p$, illustrated in Figure~\ref{fig: <b<}. Starting from (e), we deconfine the adjoint into a $\CC_p[U(N)]$ tail as in (a), then apply a sequence of dualities to reach (b). The $U(M)$ node is subsequently eliminated via the rank-reducing duality chain, leading to the expected magnetic partition function (m). Below, we focus on the steps from (a) to (b), as illustrated in Figure~\ref{fig: a to b}.

\medskip\noindent{\bf (a) $\to$ (a$'$)}.
Setting $h\equiv p-\b$, we successively apply $\CA$ for $j>h$ and $\CB$ for $j\leq h$, proceeding from $j=p$ down to $j=1$:
\begin{equation}
    \begin{tikzpicture}[thick,scale=0.7, every node/.style={scale=0.7}]%[shorten >=1pt]
  \tikzset{vertex/.style={circle,fill=white!25,minimum size=2.5em,inner sep=2pt}}
  \tikzset{every loop/.style={}}%removes arrow head from all loops.

% Theory A'
    \node[vertex] (A'n) at (0,0) [shape=circle,draw=black,minimum size=2.5em] {\scriptsize $m_{j\texttt{-}\!1}$};
    \node[vertex] (A'2n) at (1.5,0) [shape=circle,draw=black,minimum size=2.5em] {\scriptsize $m_{j}$};
    \node[vertex] (A'm) at (3,0) [shape=circle,draw=black,minimum size=2.5em] {\scriptsize $m_{j\texttt{+}1}^{(1)}\!$};
    \node[vertex] (A'cd1) at (-1.5,0) {$\cdots$};
    \node[vertex] (A'cd2) at (4.5,0) {$\cdots$};
    \node[vertex] (A'F1) at (6,1) [shape=rectangle,draw=black,minimum height=2.5em, minimum width=2.5em] {$F$};
    \node[vertex] (A'F2) at (6,-1) [shape=rectangle,draw=black,minimum height=2.5em, minimum width=2.5em] {$F$};

    \draw[-to, min distance=1cm]  (A'n) edge [out=120, in=60] node {} (A'n);
    \draw[->-=.5] ([yshift= 2pt] A'cd1.east) to ([yshift= 2pt] A'n.west);
    \draw[->-=.5] ([yshift= -2pt] A'n.west) to ([yshift= -2pt] A'cd1.east);
    \draw[->-=.5] ([yshift= 2pt] A'n.east) to ([yshift= 2pt] A'2n.west);
    \draw[->-=.5] ([yshift= -2pt] A'2n.west) to ([yshift= -2pt] A'n.east);
    \draw[->-=.5] ([yshift= 2pt] A'm.west) to ([yshift= 2pt] A'2n.east);
    \draw[->-=.5] ([yshift= -2pt] A'2n.east) to ([yshift= -2pt] A'm.west);
    \draw[->-=.5] ([yshift= 2pt] A'm.east) to ([yshift= 2pt] A'cd2.west);
    \draw[->-=.5] ([yshift= -2pt] A'cd2.west) to ([yshift= -2pt] A'm.east);
    
    \draw[->-=.5] ([yshift= 0pt] A'F1.west) to [bend right=20] ([xshift= -6pt,yshift=13pt] A'm.east);
    \draw[->-=.5] ([xshift= -6pt,yshift=-13pt] A'm.east) to [bend right=20] ([yshift= 0pt] A'F2.west);
    
    \draw[->-=.5] ([xshift= -5pt,yshift=13pt] A'2n.east) to [bend left=25] ([yshift= 5pt] A'F1.west);
    \draw[->-=.5] ([yshift= -5pt] A'F2.west) to [bend left=25] ([xshift= -5pt,yshift=-13pt] A'2n.east);

    \node at (0,1.3) {\scriptsize $2\texttt{-}2\D$};
    \node at (-0.7,0.4) {\scriptsize $\D$};
    \node at (0.8,0.4) {\scriptsize $\D$};
    \node at (2.3,0.4) {\scriptsize $1\texttt{-}\D$};
    \node at (3.8,0.4) {\scriptsize $\D$}; 
    \node at (6,0) {\scriptsize $\cdots$};

    \node at (3.6,0.9)
    {\begin{tikzpicture}
      \node [rotate=25] {\scriptsize $(j\texttt{+}2)\D\texttt{-}r$};    
    \end{tikzpicture}
    };
    \node at (2.8,1.3)
    {\begin{tikzpicture}
      \node [rotate=15] { $r \texttt{+} (p\texttt{-}j)\D$};    
    \end{tikzpicture}
    };

    \node at (0,-2+0.3) {\scriptsize $\t_{j\texttt{-}1}$};
    
    \node at (1.5,-2+0.3) {\scriptsize $t_j$};
    
    \node at (3,-2+0.3) {\scriptsize $e_{j\texttt{+}1}\texttt{-}t_{j\texttt{+}1}$};

    \node at (8,0.5) {$\CA / \CB$};
    \node at (8,0) {\LARGE $=\!\!=$};

    % Theory B'
    \node[vertex] (B'n) at (11,0) [shape=circle,draw=black,minimum size=2.5em]{\scriptsize $m_{j\texttt{-}\!1}$};
    \node[vertex] (B'2m) at (12.5,0) [shape=circle,draw=black,minimum size=2.5em] {\scriptsize $m_{j}^{(1)}\!$};
    \node[vertex] (B'm) at (14,0) [shape=circle,draw=black,minimum size=2.5em] {\scriptsize $m_{j\texttt{+}1}^{(1)}\!$};
    \node[vertex] (B'cd1) at (9.5,0) {$\cdots$};
    \node[vertex] (B'cd2) at (15.5,0) {$\cdots$};
    \node[vertex] (B'F1) at (17,1) [shape=rectangle,draw=black,minimum height=2.5em, minimum width=2.5em] {$F$};
    \node[vertex] (B'F2) at (17,-1) [shape=rectangle,draw=black,minimum height=2.5em, minimum width=2.5em] {$F$};

    \draw[->-=.5] ([yshift= 2pt] B'cd1.east) to ([yshift= 2pt] B'n.west);
    \draw[->-=.5] ([yshift= -2pt] B'n.west) to ([yshift= -2pt] B'cd1.east);
    \draw[->-=.5] ([yshift= 2pt] B'2m.west) to ([yshift= 2pt] B'n.east);
    \draw[->-=.5] ([yshift= -2pt] B'n.east) to ([yshift= -2pt] B'2m.west);
    \draw[->-=.5] ([yshift= 2pt] B'2m.east) to ([yshift= 2pt] B'm.west);
    \draw[->-=.5] ([yshift= -2pt] B'm.west) to ([yshift= -2pt] B'2m.east);
    \draw[-to, min distance=0.8cm]  (B'm) edge [out=120, in=60] node {} (B'm);
    \draw[->-=.5] ([yshift= 2pt] B'm.east) to ([yshift= 2pt] B'cd2.west);
    \draw[->-=.5] ([yshift= -2pt] B'cd2.west) to ([yshift= -2pt] B'm.east);
    \draw[->-=.5] ([yshift= 5pt] B'F1.west) to [bend right=25] ([xshift= -5pt,yshift=13pt] B'2m.east);
    \draw[->-=.5] ([xshift= -5pt,yshift=-13pt] B'2m.east) to [bend right=25] ([yshift= -5pt] B'F2.west);
    \draw[->-=.5] ([xshift= 3pt,yshift=0pt] B'n.north) to [bend left=20] ([yshift= 10pt] B'F1.west);
    \draw[->-=.5] ([yshift= -10pt] B'F2.west) to [bend left=20] ([xshift= 3pt,yshift=0pt] B'n.south);
    
    \draw[->-=.6] ([xshift= 7pt] B'F2.north) to ([xshift= 7pt] B'F1.south);

    \node at (-0.7+11,0.4) {\scriptsize $\D$};
    \node at (0.8+11,0.4) {\scriptsize $1\texttt{-}\D$};
    \node at (2.3+11,0.4) {\scriptsize $\D$};
    \node at (3.8+11,0.4) {\scriptsize $\D$};
    \node at (3.8+11,0.9) {\scriptsize $2\texttt{-}2\D$};
    \node at (18.3,0) {\scriptsize $2r \texttt{+} 2(p\texttt{-}j)\D$};
    \node at (17-0.1,0) {\scriptsize $\cdots$}; 
    
    \node at (17,-2) {$+\big\{C_j - 2t_{j} \big\}$};
    \node at (17.3,-2.5) {$+\big\{C_j \big\}_{\text{if } j>h}$};
    
    \node at (4.15+9,0.9)
    {\begin{tikzpicture}
      \node [rotate=29] { $(j\texttt{+}1)\D\texttt{-}r$};    
    \end{tikzpicture}
    };
    \node at (2.8+9.3,1.3)
    {\begin{tikzpicture}
      \node [rotate=15] { $r \texttt{+} (p\texttt{-}j\texttt{+}1)\D$};    
    \end{tikzpicture}
    };

    \node at (0+11,-2) {\scriptsize $t_{j\texttt{-}1}$};
    
    \node at (1.5+11,-2) {\scriptsize $e_j\texttt{-}t_j$};

    \node at (3+11,-2) {\scriptsize $\t_{j\texttt{+}1}^{(1)}$};

\end{tikzpicture}
\label{eq:outward-duality}
\end{equation}
The gauge rank $m^{(1)}_j$ is inductively computed as
\begin{align} \label{eq:outward-ranks}
    m^{(1)}_j &= m_{j-1} + m^{(1)}_{j+1} +F - m_j -e_j  \nn\\
    &= m_{j-1} -N + (p-j+1)F - \max (0,h-j+1) \;, 
\end{align}
where
\begin{equation}
    e_j = \begin{cases}
    1, & j\le h,\\
    0, & j>h,
    \end{cases}
    \qquad 1\le j\le p\;.
\end{equation}
Thus, $m^{(1)}_1=\WN$ and $m^{(1)}_2=n+\WN-F+1$, while the FI-parameters are determined recursively by
\begin{align}
    &t_j = \tau_j +t_{j+1} -e_{j+1}\Delta = \t_{j,p} -\Delta\max(0,h-j)  \;, \nn\\
    &\tau^{(1)}_j \equiv \tau_{j-1} + (e_j-e_{j-1})(1-\Delta) 
    \;,\;\;\; 2\leq j\leq p
    \,,
\end{align}
so that the final FI-parameters read
\begin{align}\label{eq:fi-first}
    \tau^{(1)}_1 &= 1-t_1 = 1- \tau -(\beta-\alpha) \Delta \;, \nn\\
    \tau^{(1)}_j &=\tau_{j-1} -(1-\D)\delta_{j,h+1} \;, \quad 2\leq j \leq p \;.
\end{align}
The R-charge $C_j$ of the positively charged monopole at the $j$-th node of the left-hand-side in \eqref{eq:outward-duality} is
\begin{equation}
 C_j=m_{j-1}(1-\D)+m^{(1)}_{j+1}\D +F\bigl(1-r-(p-j)\D\bigr)-m_j+1+t_j \;.
\end{equation}
Using an identity $m_{j-1}-2m_j+m_{j+1}=\delta_{j,p-\alpha}$ for $1\leq j<p$ gives rise to
\begin{equation}
    C_j=2-2\max(0,j-h)\D 
    \;,
\end{equation}
which confirms the R-charge 2 condition at the sequential $\CB$ steps for $j\leq h$.

Let us collect the generated singlets. We begin with singlets of charges $2i \Delta$ for $i=1,\cdots \min(\alpha,p-\alpha)$ by deconfining the electric theory. 
Both $\CA$ and $\CB$ generate singlet of R-charge $C_j-2t_j$ from the negative monopoles, while $\CA$ additionally yields $C_j$ from the positively charged monopole, which pair up to cancel the initial singlets. We then end up with
\begin{align}
    &\{2i\Delta\}_{i=\beta+1}^{\min(\alpha,p-\alpha)} \;, \qquad \mathrm{for} \quad \beta < \min(\alpha,p-\alpha) \;,\nn\\ 
    &\{2-2i\Delta\}_{i=\min(\alpha,p-\alpha)+1}^\beta \;, \qquad \mathrm{for} \quad \beta > \min(\alpha,p-\alpha) \;,
\end{align}
and
\begin{equation}\label{eq:first-singlet2}
    \{F(1-r) -2\Delta(N-1-i)-\t \}_{i=0}^{p-1} \;.
\end{equation}

\medskip
\noindent{\bf (a$'$) $\to$ (b$'$)}.
Deconfining the adjoint at the $U(m_2^{(1)})$ node via $\CB$, highlighted in red in Figure~\ref{fig: a to b}, introduces a $U(m_2^{(1)}-1)$ node with FI-parameter $\t^*=m_2^{(1)}(1-\D)$. The FI-parameter of the original $U(m_2^{(1)})$ node then becomes
\begin{equation}
  t^{(2)}_2=\tau^{(1)}_2-(m^{(1)}_2-1)(1-\D)
  \,,
  \label{eq:return-fi-two}
\end{equation}
and a singlet of R-charge $2\tau_*$ appears. We then successively apply $\CB$ for $j\leq h$ and $\CA$ for $j>h$, proceeding from left to right ($j=2,\ldots,p$). We define $m^{(2)}_1=m^{(1)}_2-1$.

The node to be dualized has left and right arrows of R-charges $1-\D$ and $\D$, respectively, and arrows of R-charge $(j-1)\D$ connected to the $U(\WN)$ node. As it is dualized, the meson-bridge moves the latter arrows to the adjacent right node, while complementary meson pairs cancel the existing arrows. The new ranks $m_j^{(2)}$ are then determined recursively as
\begin{align}
  m^{(2)}_j&=\WN+m^{(2)}_{j-1}+m^{(1)}_{j+1}-m^{(1)}_j-e_j \nn\\
  &=m_j+j(\WN-F),\qquad 2\leq j\leq p .
  \label{eq:return-ranks}
\end{align}
In particular, $m_p^{(2)}=M$, and the ranks $m_j^{(2)}$ match those of the $\CC_p[U(M)]$ tail. We now show that the resulting tail indeed has the form of $\CC_p[U(M)]$, as highlighted in blue in Figure~\ref{fig: a to b}.

Let \(t_j^{(2)}\) be the FI-parameter of the \(j\)-th node before dualization. Its value at \(j=2\) is given by \eqref{eq:return-fi-two}. The shift from the preceding move gives
\begin{equation}
 t^{(2)}_j=\tau^{(1)}_j+t^{(2)}_{j-1}-e_{j-1}\D\;, \qquad 3\leq j\leq p \;.
 \label{eq:return-fi-rec}
\end{equation}
With (\ref{eq:fi-first}), we obtain for $ 2\leq j\leq p$
\begin{equation}
 t^{(2)}_j=\sum_{k=1}^{j-1}\tau_k -(m^{(1)}_2-1)(1-\D) -(1-\D)(1-e_j) -\D\max\bigl(0,\min(j-1,h)-1\bigr) \;.
 \label{eq:return-fi-closed}
\end{equation}
For $2\leq j<p$, the left, $j$-th, and right FI-parameters change as
\begin{align}
 \begin{cases}
 \tau_*,&j=2,\\
 e_{j-1}-t^{(2)}_{j-1},&j\geq3
 \end{cases}
 &\rightarrow \tau^{(2)}_{j-1}
 =\begin{cases}
 \tau_*+t^{(2)}_2-e_2(1-\D),&j=2\;,\\
 e_{j-1}-t^{(2)}_{j-1}+t^{(2)}_j-e_j(1-\D),&j\geq3 \;,
 \end{cases} \nn\\
 t^{(2)}_j&\rightarrow e_j-t^{(2)}_j\;,\nn \\
 \tau^{(1)}_{j+1}&\rightarrow t^{(2)}_{j+1}=\tau^{(1)}_{j+1}+t^{(2)}_j-e_j\D \;.
 \label{eq:return-local-fi}
\end{align}
The FI-parameter of the $U(\WN)$ node also changes at each step, taking the following values before and after the $j$-th dualization, respectively:
\begin{equation}
 \tau^{(1)}_1+\sum_{k=2}^{j-1}
 \bigl(t^{(2)}_k-e_k(k-1)\D\bigr)
 \rightarrow
 \tau^{(1)}_1+\sum_{k=2}^{j}
       \bigl(t^{(2)}_k-e_k(k-1)\D\bigr).
 \label{eq:return-outside-running}
\end{equation}
At $j=p$, the right node is absent, while the other shifts still
apply.  In particular, $e_p=0$ and the last node has the final FI-parameter $\tau^{(2)}_p=-t^{(2)}_p$.
Accordingly, after the sequence of the dualities, FI-parameters of the new linear quiver tail read
\begin{equation}
 \tau^{(2)}_j=\tau_j\quad(1\leq j<p),\qquad
 \tau^{(2)}_p=\WN(1-2\D)-[M-1]_p\D .
 \label{eq:final-tail-fi}
\end{equation}
Summing \eqref{eq:return-outside-running}, the FI-parameter of the left-most $U(\WN)$ node becomes
\begin{align}
 \tau^{(1)}_1+\sum_{j=2}^{p}
       \bigl(t^{(2)}_j-e_j(j-1)\D\bigr)
 &=-\tau+
   \bigl(2p-2\beta-(p-1)\alpha
          -p(p-1)(m^{(1)}_2-1)\bigr)\D \notag\\
 &=\widetilde\tau-M(1-2\D).
\end{align}

Right before dualizing the $j$-th node, its positive monopole carries R-charge
\begin{equation}
 D_j=\WN\bigl(1-(j-1)\D\bigr) +m^{(2)}_{j-1}\D+m^{(1)}_{j+1}(1-\D) -m^{(1)}_j+1+t^{(2)}_j 
 \,.
 \label{eq:return-monopole}
\end{equation}
Combined with \eqref{eq:outward-ranks}, \eqref{eq:return-ranks}, and \eqref{eq:return-fi-closed}, it becomes
\begin{equation}
 D_j=
 \begin{cases}
  2,&2\leq j\leq h,\\
  2(j-h)\D,&h<j\leq p \,,
 \end{cases}
\end{equation}
which confirms the R-charge 2 condition of the $\CB$ dualizations.

We now collect the singlet contributions. The $\CB$-deconfinement produces a singlet of R-charge $2\tau_*$, while each backward $\CB$ contributes one of R-charge $2-2t_j^{(2)}$. Each $\CA$ generates two singlets of R-charges $2(j-h)\Delta$ and $2(j-h)\Delta-2t_j^{(2)}$, where the former cancels the singlet of R-charge $2-2(j-h)\Delta$. The remaining singlets are thus
\begin{equation}
    \{2\Delta j\}_{j=1}^{\min(\alpha,p-\alpha)} \;, \ \{2\Delta j\}_{j=M+1}^{M+p} \;, \ \{F(1-r) -2\Delta(N-1-i) - \t \}_{i=0}^{p-1} \;,
    \label{eq: last sing}
\end{equation}
where the second set comes from the negatively charged monopole.

\medskip\noindent{\bf (b$'$) $\to$ (b) }.
We finally confine the $\CC_p[U(M)]$ tail. As a result, the FI parameter of the $U(M)$ node becomes $\WN(1-2\D)$ and the first set of singlets in \eqref{eq: last sing} disappears, yielding quiver~(b) in Figure~\ref{fig: <b<}.
Eliminating the $U(M)$ node via the rank-reducing duality chain then leaves a $U(\WN)$ node with FI parameter $\tilde{\t}$ and produces additional singlets $\{2\D j\}_{j=1-M}^M$, which cancel against the second set in \eqref{eq: last sing}.
The remaining singlet contributions are
\begin{equation}
    \{F(1-r) -2\Delta(N-1-i)-\t \}_{i=0}^{p-1} \,.
\end{equation}
We thus arrive at the expected magnetic partition function (m) in Figure~\ref{fig: <b<}, completing the derivation for $0<\b<p$.

\subsubsection{For \texorpdfstring{$p<\b\leq 2p$}{p<b=<2p}}
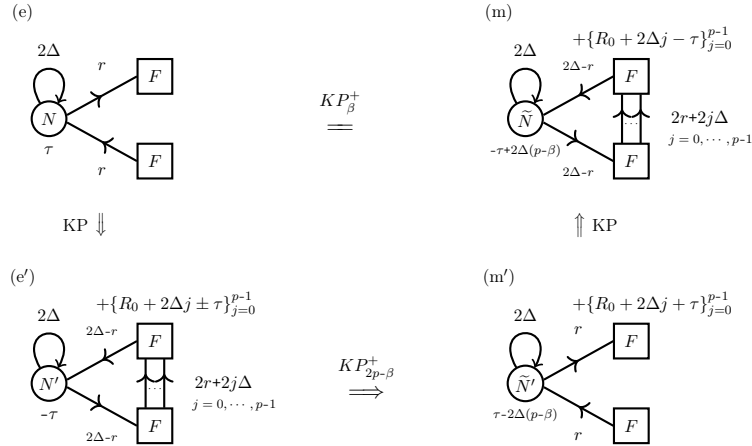
\begin{figure}[thbp]
\centering
\begin{tikzpicture}[thick,scale=0.7, every node/.style={scale=0.6}]%[shorten >=1pt]%[shorten >=1pt]
  \tikzset{vertex/.style={circle,fill=white!25,minimum size=12pt,inner sep=2pt}}
  \tikzset{every loop/.style={}}%removes arrow head from all loops.
    
    % LHS Theory
    \node[vertex] (lkmNc) at (14-9,+0) [shape=circle,draw=black,minimum size=2em] {$N'$};
    \node[vertex] (lkmNf1) at (16-9,0.8+0) [shape=rectangle,draw=black,minimum height=2em, minimum width=2em] {$F$};
    \node[vertex] (lkmNf2) at (16-9,-0.8+0) [shape=rectangle,draw=black,minimum height=2em, minimum width=2em] {$F$};

    \draw[-to, min distance=1cm]  (lkmNc) edge [out=120, in=60] node {} (lkmNc);
    \draw[->-=.5] ([yshift= 0pt] lkmNf1.west) to ([yshift= 2pt] lkmNc.east);
    \draw[->-=.5] ([yshift= -2pt] lkmNc.east) to ([yshift= 0pt] lkmNf2.west);
    \draw[->-=.7] ([xshift= -5pt] lkmNf2.north) to ([xshift= -5pt] lkmNf1.south);
    \draw[->-=.7] ([xshift= 5pt] lkmNf2.north) to ([xshift= 5pt] lkmNf1.south);

    \node at (14-0.5-9,2+0) {(e$'$)};
    
    \node at (16-0.1-9,-0.1+0) {\tiny$\cdot$};
    \node at (16-9,-0.1+0) {\tiny$\cdot$};
    \node at (16+0.1-9,-0.1+0) {\tiny$\cdot$};
    \node at (14-9,1.3+0) {$2\D$};
    \node at (15-9,1+0) {\scriptsize $2\D\texttt{-}r$};
    \node at (15-9,-1+0) {\scriptsize $2\D\texttt{-}r$};
    \node at (14-9,-0.6+0) { $\texttt{-}\t$};

    \node at (17.3-9,0+0) {$2r \texttt{+} 2j\D$};
    \node at (17.5-9,-0.4+0) {\scriptsize $j=0,\cdots,p\texttt{-}1$};

    \node at (16.4-9,1.5+0) {$+\{R_0 + 2\D j \pm \t\}_{j=0}^{p\texttt{-}1}$};
    
    % RHS Theory
    \node[vertex] (rkNc) at (14,+0) [shape=circle,draw=black,minimum size=2em] {$\WN'$};
    \node[vertex] (rkeNf1) at (16,0.8+0) [shape=rectangle,draw=black,minimum height=2em, minimum width=2em] {$F$};
    \node[vertex] (rkeNf2) at (16,-0.8+0) [shape=rectangle,draw=black,minimum height=2em, minimum width=2em] {$F$};

    \draw[-to, min distance=1cm]  (rkNc) edge [out=120, in=60] node {} (rkNc);
    \draw[->-=.5] ([yshift= 2pt] rkNc.east) to ([yshift= 0pt] rkeNf1.west);
    \draw[->-=.5] ([yshift= 0pt] rkeNf2.west) to ([yshift= -2pt] rkNc.east);

    \node at (14-0.5,2+0) {(m$'$)};

    \node at (14,1.3+0) {$2\D$};
    \node at (14,-0.6+0) {\scriptsize $\t \texttt{-} 2\D(p\texttt{-}\b)$};
    \node at (15,1+0) {$r$};
    \node at (15,-1+0) {$r$};
    \node at (16.4,1.5+0) {$+\{R_0 + 2\D j + \t\}_{j=0}^{p\texttt{-}1}$};
    
    % vertical arrows
    \node at (7-1,+3) 
    {\begin{tikzpicture}
      \node [rotate=90] {\LARGE\textcolor{black}{$\Longleftarrow$}};    
    \end{tikzpicture}
    };
    \node at (7-1.5,+3) {KP};

    \node at (15,+3) 
    {\begin{tikzpicture}
      \node [rotate=-90] {\LARGE\textcolor{black}{$\Longleftarrow$}};    
    \end{tikzpicture}
    };
    \node at (15+0.5,+3) {KP};
    
    % electric
    \node[vertex] (kNc) at (5,+5) [shape=circle,draw=black,minimum size=2em] {$N$};
    \node[vertex] (keNf1) at (7,0.8+5) [shape=rectangle,draw=black,minimum height=2em, minimum width=2em] {$F$};
    \node[vertex] (keNf2) at (7,-0.8+5) [shape=rectangle,draw=black,minimum height=2em, minimum width=2em] {$F$};

    \draw[-to, min distance=1cm]  (kNc) edge [out=120, in=60] node {} (kNc);
    \draw[->-=.5] ([yshift= 2pt] kNc.east) to ([yshift= 0pt] keNf1.west);
    \draw[->-=.5] ([yshift= 0pt] keNf2.west) to ([yshift= -2pt] kNc.east);

    \node at (5-0.5,2+5) {(e)};

    \node at (5,1.3+5) {$2\D$};
    \node at (5,-0.6+5) {$\t$};
    \node at (6,1+5) {$r$};
    \node at (6,-1+5) {$r$};

    % magnetic
    \node[vertex] (kmNc) at (14,+5) [shape=circle,draw=black,minimum size=2em] {$\WN$};
    \node[vertex] (kmNf1) at (16,0.8+5) [shape=rectangle,draw=black,minimum height=2em, minimum width=2em] {$F$};
    \node[vertex] (kmNf2) at (16,-0.8+5) [shape=rectangle,draw=black,minimum height=2em, minimum width=2em] {$F$};

    \draw[-to, min distance=1cm]  (kmNc) edge [out=120, in=60] node {} (kmNc);
    \draw[->-=.5] ([yshift= 0pt] kmNf1.west) to ([yshift= 2pt] kmNc.east);
    \draw[->-=.5] ([yshift= -2pt] kmNc.east) to ([yshift= 0pt] kmNf2.west);
    \draw[->-=.7] ([xshift= -5pt] kmNf2.north) to ([xshift= -5pt] kmNf1.south);
    \draw[->-=.7] ([xshift= 5pt] kmNf2.north) to ([xshift= 5pt] kmNf1.south);

    \node at (14-0.5,2+5) {(m)};
    
    \node at (16-0.1,-0.1+5) {\tiny$\cdot$};
    \node at (16,-0.1+5) {\tiny$\cdot$};
    \node at (16+0.1,-0.1+5) {\tiny$\cdot$};
    \node at (14,1.3+5) {$2\D$};
    \node at (15,1+5) {\scriptsize $2\D\texttt{-}r$};
    \node at (15,-1+5) {\scriptsize $2\D\texttt{-}r$};
    \node at (14,-0.6+5) {\scriptsize $\texttt{-}\t \texttt{+}2\D(p\texttt{-}\b) $};

    \node at (17.3,0+5) {$2r \texttt{+} 2j\D$};
    \node at (17.5,-0.4+5) {\scriptsize $j=0,\cdots,p\texttt{-}1$};

    \node at (16.4,1.5+5) {$+\{R_0 + 2\D j - \t\}_{j=0}^{p\texttt{-}1}$};

    \node at (10.5,0.3+5) {$KP_\b^+$};
    \node at (10.5,-0.2+5) {\Large $=\!\!=$};

    \node at (11,0.3) {$KP_{2p\texttt{-}\b}^+$};
    \node at (11,-0.2) {\Large $\Longrightarrow$};
    
\end{tikzpicture}
\caption{\label{fig: big b} Successive dualization for deriving the $KP_\b^+$ identity for $p<\b\leq 2p$.}
\end{figure}
Consider (e) in Figure \ref{fig: big b} where $\t$ is fixed by the monopole superpotential term
\begin{align}
    \underbrace{F(1-r)-2\D(N-1)}_{= R_0}+2\D \b + \t = 2
    \;\;,\;\;\;\;
    \text{for}\;
    p<\b\leq 2p.
\end{align}
Applying the original KP duality gives rise to (e$'$) with $N' = pF-N$, in which frame, the $(2p-\b)$-th dressed monopole operator has R-charge
\begin{align}
    F(1-2\D + r) - 2\D(N'-1) + 2\D (2p-\b) + (-\t) = 2
    \,.
\end{align}
Since $0\leq 2p-\b<p$, we can apply the previously established $KP_{2p-\b}^+$ identity to obtain quiver (m$'$) with $\WN'=pF-N'-p+(2p-\b)=N+p-\b$. The resulting singlet contributions cancel the preceding ones $+\{R_0+2\D j-\t\}_{j=0}^{p-1}$. A final application of the original KP duality then gives (m) with the desired dual rank $\WN=pF-\WN'=pF-N-p+\b$. The newly generated singlet contributions also cancel the existing ones, leaving precisely the expected contributions. We thus demonstrate the $KP_\b^+$ duality at the level of ellipsoid partition functions using only Aharony and BBP dualities.

\section*{Acknowledgements}
We thank Chiung Hwang and Yuanyuan Fang for discussions and related collaboration. We acknowledge the use of ChatGPT/Codex in the development of {\tt Quiver Lab}. The work of SK and JL is supported by the National Research Foundation of Korea (NRF) Grant RS-2024-00405629. SK is supported by the NRF Grant RS-2023-NR076601. JL is supported by the NRF Grant RS-2026-25482546.

\appendix
\section{Alternative partition function derivation of s-confinement of \texorpdfstring{$\CC_p[U(N)]$}{Dp[U(N)]}}\label{app: conf derivation}
We provide an alternative proof of the $\CC_p[U(N)]$ confinement by using the BBP duality. We first treat the $N\leq p$ case, then consider $p<N$ cases with mathematical induction together with the rank-reducing duality chain.

\medskip\noindent{\bf $\bullet$\; The $N\leq p$ case}.
If $N\leq p$, the ellipsoid partition function of $\CC_p[U(N)]$ reads ($\D \equiv \frac{1}{p+1}$)
\begin{equation}
    \begin{tikzpicture}[thick,scale=0.8, every node/.style={scale=0.8}]%[shorten >=1pt]
      \tikzset{vertex/.style={circle,fill=white!25,minimum size=12pt,inner sep=2pt}}
      \tikzset{every loop/.style={thick}}%removes arrow head from all loops.
        % Theory A
    \node[vertex] (A1) at (0,0) [shape=circle,draw=black,minimum size=2em] {\scriptsize $1$};
    \node[vertex] (A2) at (1.5,0) [shape=circle,draw=black,minimum size=2em] {\scriptsize $2$};
    \node[vertex] (Acd) at (3,0)  {$\cdots$};
    \node[vertex] (Ap2) at (4.5,0) [shape=circle,draw=black,minimum size=2em] { \scriptsize$N\texttt{-}1$};
    \node[vertex] (Ap1) at (6,0) [shape=rectangle,draw=black,minimum size=2em] {\scriptsize $N$};

    \draw[-to, min distance=0.8cm]  (A1) edge [out=120, in=60] node {} (A1);
    \draw[-to, min distance=0.8cm]  (A2) edge [out=120, in=60] node {} (A2);
    \draw[-to, min distance=0.8cm]  (Ap2) edge [out=120, in=60] node {} (Ap2);
    \draw[->-=.5] ([yshift= 2pt] A1.east) to ([yshift= 2pt] A2.west);
    \draw[->-=.5] ([yshift= -2pt] A2.west) to ([yshift= -2pt] A1.east);
    \draw[->-=.5] ([yshift= 2pt] A2.east) to ([yshift= 2pt] Acd.west);
    \draw[->-=.5] ([yshift= -2pt] Acd.west) to ([yshift= -2pt] A2.east);
    \draw[->-=.5] ([yshift= 2pt] Acd.east) to ([yshift= 2pt] Ap2.west);
    \draw[->-=.5] ([yshift= -2pt] Ap2.west) to ([yshift= -2pt] Acd.east);
    \draw[->-=.5] ([yshift= 2pt] Ap2.east) to ([yshift= 2pt] Ap1.west);
    \draw[->-=.5] ([yshift= -2pt] Ap1.west) to ([yshift= -2pt] Ap2.east);

    \node at (0.7,0.4) {\scriptsize $\D$};
    \node at (2.2,0.4) {\scriptsize $\D$};
    \node at (3.7,0.4) {\scriptsize $\D$};
    \node at (5.2,0.4) {\scriptsize $\D$};
    
    \node at (0,1.1) {\scriptsize $2\texttt{-}2\D$};
    \node at (1.5,1.1) {\scriptsize $2\texttt{-}2\D$};
    \node at (4.5,1.1) {\scriptsize $2\texttt{-}2\D$};
    
    \node at (0,-0.7) {\scriptsize $2\D$};
    \node at (1.5,-0.7) {\scriptsize $2\D$};
    \node at (4.5,-0.7) {\scriptsize $2\D$};
        
    \end{tikzpicture}
    \label{eq: small N}
\end{equation}
This is the $T[U(N)]$ theory~\cite{Gaiotto:2008ak} with tuned FI-parameters. The adjoint at the $U(1)$ node is a singlet of R-charge $2-2\D$, allowing us to apply a sequence of BBP dualities starting from this node as follows:
\begin{equation}
\begin{tikzpicture}[thick,scale=0.8, every node/.style={scale=0.8}]%[shorten >=1pt]
      \tikzset{vertex/.style={circle,fill=white!25,minimum size=12pt,inner sep=2pt}}
      \tikzset{every loop/.style={}}%removes arrow head from all loops.
        \node[vertex] (n1) at (1-0.3,0) [shape=circle,draw=black,minimum size=2em] {\scriptsize $j$};
        \node[vertex] (n2) at (2.5-0.3,0) [shape=circle,draw=black,minimum height=2em, minimum width=2em] {\scriptsize $j\!+\!1$};
        \node[vertex] (n3) at (4-0.3,0) {$\cdots$};
    
        \draw[-to, min distance=1cm]  (n2) edge [out=120, in=60] node {} (n2);
        \draw[->-=.5] ([yshift= 2pt] n1.east) to ([yshift= 2pt] n2.west);
        \draw[->-=.5] ([yshift= -2pt] n2.west) to ([yshift= -2pt] n1.east);
        \draw[->-=.5] ([yshift= 2pt] n2.east) to ([yshift= 2pt] n3.west);
        \draw[->-=.5] ([yshift= -2pt] n3.west) to ([yshift= -2pt] n2.east);

        \node at (1.75-0.3,0.4) {\scriptsize $\D$};
        \node at (3.25-0.3,0.4) {\scriptsize $\D$};
        \node at (2.5-0.3,1.2) {\scriptsize $2\texttt{-}2\D$};
        
        \node at (1-0.3,-0.7) {\tiny $(j\texttt{+}1)\D$};
        \node at (2.5-0.3,-0.7) {\tiny $2\D$};

        \node at (5.3,0.5) {\scriptsize $\CB$};
        \node at (5.3,0) {$\Rightarrow$};
        
        \node[vertex] (N1) at (7.6,0) [shape=circle,draw=black,minimum size=2em] {\scriptsize $j\!+\!1$};
        \node[vertex] (N2) at (9.1,0) {$\cdots$};

        \draw[->-=.5] ([yshift= 2pt] N1.east) to ([yshift= 2pt] N2.west);
        \draw[->-=.5] ([yshift= -2pt] N2.west) to ([yshift= -2pt] N1.east);
        
        \node at (8.35,0.4) {\scriptsize $\D$};
        
        \node at (7.6,-0.7) {\tiny $(j\texttt{+}2)\D$};
    
        \node at (9.5,1) {\scriptsize $+\{ 2\D(p-j) \}$};
        
    \end{tikzpicture}
\end{equation}
for $j=1,\ldots,N-1$, leading to the confinement of $\CC_p[U(N)]$ for $N\leq p$:
\begin{equation}
    \begin{tikzpicture}[thick,scale=0.8, every node/.style={scale=0.8}]%[shorten >=1pt]
      \tikzset{vertex/.style={circle,fill=white!25,minimum size=12pt,inner sep=2pt}}
      \tikzset{every loop/.style={}}%removes arrow head from all loops.
        \node[vertex] (N) at (0,0) [shape=rectangle,draw=black,minimum height=2em, minimum width=2em] {$N$};
    
        \draw[-to, min distance=1cm]  (N) edge [out=120, in=60] node {} (N);
    
        \node at (-4,0) {$\CC_p[U(N\leq p)]$};
        \node at (-4,0.7) {$+\{2\D j\}_{j=1}^{[N]_p}$};
        
        \node at (-1.8,0.2) {$\xRightarrow{\text{conf}}$};
        
        \node at (0,1.2) {\tiny $2\D$};
    \end{tikzpicture}
\end{equation}
We flipped the singlet contribution to the left-hand-side.

\medskip\noindent{\bf $\bullet\;$ Induction for $p < N$}. Now, let us consider $\CC_p[U(N+p)]$ for some $N$:
\begin{equation}
    \begin{tikzpicture}[thick,scale=0.8, every node/.style={scale=0.8}]%[shorten >=1pt]
      \tikzset{vertex/.style={circle,fill=white!25,minimum size=12pt,inner sep=2pt}}
      \tikzset{every loop/.style={}}%removes arrow head from all loops.
        % Theory A
    \node[vertex] (A1) at (0,0) [shape=circle,draw=black,minimum size=2em] {\scriptsize $m_1$};
    \node[vertex] (A2) at (1.5,0) [shape=circle,draw=black,minimum size=2em] {\scriptsize $m_2$};
    \node[vertex] (Acd) at (3,0)  {$\cdots$};
    \node[vertex] (Ap2) at (4.5,0) [shape=circle,draw=black,minimum size=2em] { \scriptsize$m_{p\texttt{-}\!1}\!$};
    \node[vertex] (Ap1) at (6,0) [shape=rectangle,draw=black,minimum size=2em] {\scriptsize $N\texttt{+}p$};

    \draw[-to, min distance=0.8cm]  (A1) edge [out=120, in=60] node {} (A1);
    \draw[-to, min distance=0.8cm]  (A2) edge [out=120, in=60] node {} (A2);
    \draw[-to, min distance=0.8cm]  (Ap2) edge [out=120, in=60] node {} (Ap2);
    \draw[->-=.5] ([yshift= 2pt] A1.east) to ([yshift= 2pt] A2.west);
    \draw[->-=.5] ([yshift= -2pt] A2.west) to ([yshift= -2pt] A1.east);
    \draw[->-=.5] ([yshift= 2pt] A2.east) to ([yshift= 2pt] Acd.west);
    \draw[->-=.5] ([yshift= -2pt] Acd.west) to ([yshift= -2pt] A2.east);
    \draw[->-=.5] ([yshift= 2pt] Acd.east) to ([yshift= 2pt] Ap2.west);
    \draw[->-=.5] ([yshift= -2pt] Ap2.west) to ([yshift= -2pt] Acd.east);
    \draw[->-=.5] ([yshift= 2pt] Ap2.east) to ([yshift= 2pt] Ap1.west);
    \draw[->-=.5] ([yshift= -2pt] Ap1.west) to ([yshift= -2pt] Ap2.east);

    \node at (0.7,0.4) {\scriptsize $\D$};
    \node at (2.2,0.4) {\scriptsize $\D$};
    \node at (3.7,0.4) {\scriptsize $\D$};
    \node at (5.2,0.4) {\scriptsize $\D$};
    
    \node at (0,1.1) {\scriptsize $2\texttt{-}2\D$};
    \node at (1.5,1.1) {\scriptsize $2\texttt{-}2\D$};
    \node at (4.5,1.1) {\scriptsize $2\texttt{-}2\D$};
    
    \node at (0,-0.7) {\scriptsize $\t_1$};
    \node at (1.5,-0.7) {\scriptsize $\t_2$};
    \node at (4.5,-0.7) {\scriptsize $\t_{p\texttt{-}1}$};
        
    \end{tikzpicture}
\end{equation}
where $m_j = j \lfloor N / p \rfloor + j + \a_j$, $\a_j =\max(0,j-p+\a)$, $\a = N\!\!\mod p$ and $\t_j = (2-p\, \d_{j,p-\a})\D$. We apply successive BBP dualities 
\begin{equation}
    \begin{tikzpicture}[thick,scale=0.8, every node/.style={scale=0.8}]%[shorten >=1pt]
      \tikzset{vertex/.style={circle,fill=white!25,minimum size=12pt,inner sep=2pt}}
      \tikzset{every loop/.style={}}%removes arrow head from all loops.
        \node[vertex] (n1) at (0,0) [shape=circle,draw=black,minimum size=2em] {\scriptsize $n_j$};
        \node[vertex] (n2) at (1.5,0) [shape=circle,draw=black,minimum height=2em, minimum width=2em] {\scriptsize $m_j$};
        \node[vertex] (n3) at (3,0) [shape=circle,draw=black,minimum height=2em, minimum width=2em] {\scriptsize $m_{j\texttt{+}1\!\!}$};
    
        \draw[-to, min distance=1cm]  (n3) edge [out=120, in=60] node {} (n3);
        \draw[->-=.5] ([yshift= 2pt] n1.east) to ([yshift= 2pt] n2.west);
        \draw[->-=.5] ([yshift= -2pt] n2.west) to ([yshift= -2pt] n1.east);
        \draw[->-=.5] ([yshift= 2pt] n2.east) to ([yshift= 2pt] n3.west);
        \draw[->-=.5] ([yshift= -2pt] n3.west) to ([yshift= -2pt] n2.east);
    
        \node at (-0.7,0) {$\cdots$};
        \node at (3.7,0) {$\cdots$};
        \node at (0.75,0.4) {\scriptsize $1\texttt{-}\D$};
        \node at (2.25,0.4) {\scriptsize $\D$};
        \node at (3,1.2) {\scriptsize $2\texttt{-}2\D$};
        
        \node at (0,-0.7) {\tiny $\texttt{-}\k_{j\texttt{-}1}\texttt{+}1$};
        \node at (1.5,-0.7) {\tiny $\k_j$};
        \node at (3,-0.7) {\tiny $\t_{j\texttt{+}1}$};
    
        \node at (5.3,0.5) {\scriptsize $\CB$};
        \node at (5.3,0) {$\Rightarrow$};
    
        \node[vertex] (N1) at (7.6,0) [shape=circle,draw=black,minimum size=2em] {\scriptsize $n_j$};
        \node[vertex] (N2) at (9.1,0) [shape=circle,draw=black,minimum height=2em, minimum width=2em] {\scriptsize $n_{j\texttt{+}1}\!\!$};
        \node[vertex] (N3) at (10.6,0) [shape=circle,draw=black,minimum height=2em, minimum width=2em] {\scriptsize $m_{j\texttt{+}1}\!\!$};
    
        \draw[-to, min distance=1cm]  (N1) edge [out=120, in=60] node {} (N1);
        \draw[->-=.5] ([yshift= 2pt] N1.east) to ([yshift= 2pt] N2.west);
        \draw[->-=.5] ([yshift= -2pt] N2.west) to ([yshift= -2pt] N1.east);
        \draw[->-=.5] ([yshift= 2pt] N2.east) to ([yshift= 2pt] N3.west);
        \draw[->-=.5] ([yshift= -2pt] N3.west) to ([yshift= -2pt] N2.east);
    
        \node at (6.9,0) {$\cdots$};
        \node at (11.3,0) {$\cdots$};
        \node at (7.6,1.2) {\scriptsize $2\texttt{-}2\D$};
        \node at (8.35,0.4) {\scriptsize $\D$};
        \node at (9.85,0.4) {\scriptsize $1\texttt{-}\D$};
        
        \node at (7.6,-0.7) {\tiny $\t_j$};
        \node at (9.1,-0.7) {\tiny $\texttt{-}\k_j\texttt{+}1$};
        \node at (10.6,-0.7) {\tiny $\k_{j\texttt{+}1}$};
    
        \node at (10.5,1) {\scriptsize $+\big\{2\texttt{-}2\k_j \big\}$};
        
    \end{tikzpicture}
\end{equation}
for $j=0,\cdots, p-1$ with $n_j = m_j - j$. Note that $j=0$ corresponds to the BBP-deconfinement as the first step. The FI-parameter $\k_j$ is inductively computed as
\begin{align}
    \k_j = 1-(p m_1+j) \D + \sum_{i=1}^j \t_i
    \,.
\end{align}
After the last BBP, the resulting tail forms $\CC_p[U(N)]$. Assuming its confinement into an adjoint of R-charge $2\D$ gives rise to
\begin{equation}
    \begin{tikzpicture}[thick,scale=0.8, every node/.style={scale=0.8}]%[shorten >=1pt]
      \tikzset{vertex/.style={circle,fill=white!25,minimum size=12pt,inner sep=2pt}}
      \tikzset{every loop/.style={}}%removes arrow head from all loops.
        % Theory A
    \node[vertex] (A1) at (0,0) [shape=circle,draw=black,minimum size=2em] {\scriptsize $n_1$};
    \node[vertex] (A2) at (1.5,0) [shape=circle,draw=black,minimum size=2em] {\scriptsize $n_2$};
    \node[vertex] (Acd) at (3,0)  {$\cdots$};
    \node[vertex] (Ap2) at (4.5,0) [shape=circle,draw=black,minimum size=2em] { \scriptsize$\!\!n_{p\texttt{-}1}\!\!$};
    \node[vertex] (Ap1) at (6,0) [shape=circle,draw=black,minimum size=2em] {\scriptsize $N$};
    \node[vertex] (AF) at (7.5,0) [shape=rectangle,draw=black,minimum height=2em, minimum width=2em] {\scriptsize $\!N\!+\!p\!$};

    \draw[-to, min distance=0.8cm]  (A1) edge [out=120, in=60] node {} (A1);
    \draw[-to, min distance=0.8cm]  (A2) edge [out=120, in=60] node {} (A2);
    \draw[-to, min distance=0.8cm]  (Ap2) edge [out=120, in=60] node {} (Ap2);
    \draw[-to, min distance=0.8cm]  (AF) edge [out=120, in=60] node {} (AF);
    \draw[->-=.5] ([yshift= 2pt] A1.east) to ([yshift= 2pt] A2.west);
    \draw[->-=.5] ([yshift= -2pt] A2.west) to ([yshift= -2pt] A1.east);
    \draw[->-=.5] ([yshift= 2pt] A2.east) to ([yshift= 2pt] Acd.west);
    \draw[->-=.5] ([yshift= -2pt] Acd.west) to ([yshift= -2pt] A2.east);
    \draw[->-=.5] ([yshift= 2pt] Acd.east) to ([yshift= 2pt] Ap2.west);
    \draw[->-=.5] ([yshift= -2pt] Ap2.west) to ([yshift= -2pt] Acd.east);
    \draw[->-=.5] ([yshift= 2pt] Ap2.east) to ([yshift= 2pt] Ap1.west);
    \draw[->-=.5] ([yshift= -2pt] Ap1.west) to ([yshift= -2pt] Ap2.east);
    \draw[->-=.5] ([yshift= 2pt] Ap1.east) to ([yshift= 2pt] AF.west);
    \draw[->-=.5] ([yshift= -2pt] AF.west) to ([yshift= -2pt] Ap1.east);

    \node at (0.7,0.4) {\scriptsize $\D$};
    \node at (2.2,0.4) {\scriptsize $\D$};
    \node at (3.7,0.4) {\scriptsize $\D$};
    \node at (5.2,0.4) {\scriptsize $\D$};
    \node at (6.7,0.4) {\scriptsize $1\texttt{-}\D$};
    
    \node at (0,1.1) {\scriptsize $2\texttt{-}2\D$};
    \node at (1.5,1.1) {\scriptsize $2\texttt{-}2\D$};
    \node at (4.5,1.1) {\scriptsize $2\texttt{-}2\D$};
    \node at (7.5,1.1) {\scriptsize $2\D$};
    
    \node at (0,-0.7) {\scriptsize $\t_1$};
    \node at (1.5,-0.7) {\scriptsize $\t_2$};
    \node at (4.5,-0.7) {\scriptsize $\t_{p\texttt{-}1}$};
    \node at (6,-0.7) {\scriptsize $1\texttt{-}\k_{p\texttt{-}1}$};

    \node at (7.5,1.5) {\scriptsize $+\{ 2-2\k_j\}_{j=0}^{p\texttt{-}1}$};

    \node at (8.5,0.3) {\scriptsize conf};
    \node at (8.5,0) {$\Rightarrow$};

    %confined
        \node[vertex] (n) at (9.5,0) [shape=circle,draw=black,minimum size=2em] {\scriptsize $N$};
        \node[vertex] (2n+1) at (11,0) [shape=rectangle,draw=black,minimum height=2em, minimum width=2em] {\scriptsize $\! N\!+\!p\!$};
    
        \draw[-to, min distance=0.8cm]  (n) edge [out=120, in=60] node {} (n);
        \draw[-to, min distance=0.8cm]  (2n+1) edge [out=120, in=60] node {} (2n+1);
        \draw[->-=.5] ([yshift= 2pt] n.east) to ([yshift= 2pt] 2n+1.west);
        \draw[->-=.5] ([yshift= -2pt] 2n+1.west) to ([yshift= -2pt] n.east);
    
        \node at (9.5,1.1) {\scriptsize $2\D$};
        \node at (10.2,0.3) {\scriptsize $1\texttt{-}\D$};
        \node at (11,1.1) {\scriptsize $2\D$};
        \node at (9.5,-0.7) {\tiny $(N\texttt{+}p)\D$};

        \node at (11,2) {\scriptsize $+\{ 2-2\k_j\}_{j=0}^{p\texttt{-}1}$};
        \node at (11,1.5) {\scriptsize $+\{ 2-2\D j\}_{j=1}^{[N]_p}$};

    \end{tikzpicture}
\end{equation}
The FI-parameter on the right-hand side satisfies the condition for the rank-reducing duality chain. Upon the disappearance of the $U(N)$ node, a singlet contribution $+\{2\D(j-N)\}_{j=1}^p$ arises, which can be rewritten as
\begin{align}
    +\big\{
    2\D(j-N)
    \big\}_{j=1}^{p}
    =
    \big\{
    2\k_j
    \big\}_{j=0}^{p-1}
    \,,
\end{align}
which cancels $+\{2-2\k_j\}_{j=0}^{p-1}$. Therefore, we arrive at
\begin{equation}
    \begin{tikzpicture}[thick,scale=0.8, every node/.style={scale=0.8}]%[shorten >=1pt]
      \tikzset{vertex/.style={circle,fill=white!25,minimum size=12pt,inner sep=2pt}}
      \tikzset{every loop/.style={}}%removes arrow head from all loops.
        \node[vertex] (N) at (0,0) [shape=rectangle,draw=black,minimum height=2em, minimum width=2em] {\scriptsize $N\!+\!p$};
    
        \draw[-to, min distance=1cm]  (N) edge [out=120, in=60] node {} (N);
    
        \node at (-4,0) {$\CC_p[U(N+ p)]$};
        \node at (-4,0.7) {$+\{2\D j\}_{j=1}^{[N]_p}$};
        
        \node at (-1.8,0.2) {$\xRightarrow{\text{conf}}$};
        
        \node at (0,1.2) {\tiny $2\D$};
    \end{tikzpicture}
\end{equation}
where we have moved the singlet contribution to the left-hand side. This establishes the confinement of $\CC_p[U(N+p)]$ assuming that of $\CC_p[U(N)]$. Since the base cases $N\leq p$ have already been verified, mathematical induction proves the confinement of $\CC_p[U(M)]$ for arbitrary positive integers $M$.

\section{Rank-reducing duality chain}\label{app: rr}
We prove the identity underlying the rank-reducing duality chain \eqref{eq: rr}. The special choices of FI-parameter $\t=\pm M(1-\D)$ lead to the following duality chain:
\begin{equation}
    \begin{tikzpicture}[thick,scale=0.8, every node/.style={scale=0.8}]%[shorten >=1pt]
\tikzset{vertex/.style={circle,fill=white!25,minimum size=12pt,inner sep=2pt}}
  \tikzset{every loop/.style={}}%removes arrow head from all loops.
    % Theory A
    \node at (-0.3,1) {\scriptsize (a)};
    \node[vertex] (AN) at (0.7,0) [shape=circle,draw=black,minimum size=2em] {$N$};
    \node[vertex] (AF) at (2.3,0) [shape=circle,draw=black,minimum height=2em, minimum width=2em] {$M$};

    \draw[-to, min distance=1cm]  (AN) edge [out=120, in=60] node {} (AN);
    \draw[->-=.5] ([yshift= 2pt] AN.east) to ([yshift= 2pt] AF.west);
    \draw[->-=.5] ([yshift= -2pt] AF.west) to ([yshift= -2pt] AN.east);

    \node at (0.7,1.2) {\scriptsize $2\d$};
    \node at (1.5,0.4) {\scriptsize $\D$};
    \node at (0.7,-0.7) {\scriptsize $\t$};
    \node at (2.3,-0.7) {\scriptsize $\k$};

    % Theory B
    \node at (-0.5,-1.5) {\scriptsize (b)};
    \node[vertex] (BN1) at (0,-2.5) [shape=circle,draw=black,minimum size=2em] {\scriptsize $N\texttt{-}1$};
    \node[vertex] (BN) at (1.5,-2.5) [shape=circle,draw=black,minimum height=2em, minimum width=2em] {$N$};
    \node[vertex] (BF) at (3,-2.5) [shape=circle,draw=black,minimum height=2em, minimum width=2em] {$M$};

    \draw[->-=.5] ([yshift= 2pt] BN1.east) to ([yshift= 2pt] BN.west);
    \draw[->-=.5] ([yshift= -2pt] BN.west) to ([yshift= -2pt] BN1.east);
    \draw[->-=.5] ([yshift= 2pt] BN.east) to ([yshift= 2pt] BF.west);
    \draw[->-=.5] ([yshift= -2pt] BF.west) to ([yshift= -2pt] BN.east);

    \node at (0.75,-2.5+0.4) {\scriptsize $\d$};
    \node at (2.2,-2.5+0.4) {\scriptsize $\D$};
    \node at (0,-2.5-0.7) {\scriptsize $\pm N\d$};
    \node at (1.5,-2.5-0.7) {\scriptsize $\t \! \mp \! \d (N\texttt{-}1)$};
    \node at (3,-2.5-0.7) {\scriptsize $\k$};
    \node at (4.3,-2.5) {\scriptsize $+\{2\d N\}$};

    % Theory C
    \node at (3.2,-4) {\scriptsize (c)};
    \node[vertex] (CN1) at (4,-5) [shape=circle,draw=black,minimum size=2em] {\scriptsize $N\texttt{-}1$};
    \node[vertex] (CF1) at (5.5,-5) [shape=circle,draw=black,minimum height=2em, minimum width=2em] {\scriptsize $M\texttt{-}1$};
    \node[vertex] (CF) at (7,-5) [shape=circle,draw=black,minimum height=2em, minimum width=2em] {$M$};

    \draw[-to, min distance=1cm]  (CN1) edge [out=115, in=65] node {} (CN1);
    \draw[-to, min distance=1cm]  (CF) edge [out=120, in=60] node {} (CF);
    \draw[->-=.5] ([yshift= 2pt] CN1.east) to ([yshift= 2pt] CF1.west);
    \draw[->-=.5] ([yshift= -2pt] CF1.west) to ([yshift= -2pt] CN1.east);
    \draw[->-=.5] ([yshift= 2pt] CF1.east) to ([yshift= 2pt] CF.west);
    \draw[->-=.5] ([yshift= -2pt] CF.west) to ([yshift= -2pt] CF1.east);

    \draw[->-=.5] ([xshift=-3pt,yshift= 8pt] CN1.east) to [bend right=-70] ([xshift= 0pt,yshift=8pt] CF.west);
    \draw[->-=.5] ([xshift= 1pt,yshift=11pt] CF.west) to [bend right=64] ([xshift=-5pt,yshift= 9pt] CN1.east);

    \node at (4.75,-5+0.4) {\scriptsize $1\texttt{-}\d$};
    \node at (6.25,-5+0.4) {\scriptsize $1\texttt{-}\D$};
    \node at (4,-5+1.2) {\scriptsize $2\d$};
     \node at (5.5,-5+1.3) {\scriptsize $\d\texttt{+}\D$};
    \node at (7,-5+1.2) {\scriptsize $2\D$};
    \node at (4,-5-0.7) {\scriptsize $\t \! \pm \! \d$};
    \node at (5.5,-5-0.7) {\scriptsize $-\t\!\pm\!\d(N\texttt{-}1)$};
    \node at (9.5,-5+0.2) {\scriptsize $+\{2N\d,2M(1\texttt{-}\D)\texttt{-}2\d(N\texttt{-}1),0\}$};
    \node at (7,-5-0.5) {\scriptsize $\k\!+\!\t$};
    \node at (7,-5-1+0.2) {\scriptsize $\mp\!\d(N\texttt{-}1)$};

    % Theory D
    \node at (7.5,-1.5) {\scriptsize (d)};
    \node[vertex] (DN) at (8,-2.5) [shape=circle,draw=black,minimum size=2em] {\scriptsize $N\texttt{-}1$};
    \node[vertex] (DN1) at (9.5,-2.5) [shape=circle,draw=black,minimum height=2em, minimum width=2em] {\scriptsize $N\texttt{-}1$};
    \node[vertex] (DF) at (11,-2.5) [shape=circle,draw=black,minimum height=2em, minimum width=2em] {$M$};

    \draw[->-=.5] ([yshift= 2pt] DN.east) to ([yshift= 2pt] DN1.west);
    \draw[->-=.5] ([yshift= -2pt] DN1.west) to ([yshift= -2pt] DN.east);
    \draw[->-=.5] ([yshift= 2pt] DN1.east) to ([yshift= 2pt] DF.west);
    \draw[->-=.5] ([yshift= -2pt] DF.west) to ([yshift= -2pt] DN1.east);

    \node at (8.75,-2.5+0.4) {\scriptsize $\d$};
    \node at (10.2,-2.5+0.4) {\scriptsize $\D$};
    \node at (8,-2.5-0.7) {\tiny $\pm \! 1 \! \pm \! \d(N\texttt{-}1)$};
    \node at (9.5,-2.4-0.7) {\tiny $\t\!\mp\!1$};
    \node at (9.5,-2.7-0.7) {\tiny $\mp\d(N\texttt{-}1)$};
    \node at (11,-2.4-0.7) {\tiny $\k \!\pm\! (1\!\texttt{-}\!\D)$};
    \node at (12.3,-2.5) {\scriptsize $+\{2\d N,0\}$};

    % Theory E
    \node at (7.2,1) {\scriptsize (e)};
    \node[vertex] (EN) at (8.2,0) [shape=circle,draw=black,minimum size=2em] {\scriptsize $N\texttt{-}1$};
    \node[vertex] (EF) at (9.8,0) [shape=circle,draw=black,minimum height=2em, minimum width=2em] {$M$};

    \draw[-to, min distance=1cm]  (EN) edge [out=120, in=60] node {} (EN);
    \draw[->-=.5] ([yshift= 2pt] EN.east) to ([yshift= 2pt] EF.west);
    \draw[->-=.5] ([yshift= -2pt] EF.west) to ([yshift= -2pt] EN.east);

    \node at (8.2,1.2) {\scriptsize $2\d$};
    \node at (9,0.4) {\scriptsize $\D$};
    \node at (8.2,-0.7) {\scriptsize $\t$};
    \node at (9.8,-0.7) {\scriptsize $\k \!\pm\! (1\texttt{-}\D)$};
    \node at (11.7,0) {\scriptsize $+\{2\d N,2\d(1\texttt{-}N)\}$};

    % Arrows
    \node at (1.5,-1.3) 
    {\begin{tikzpicture}
      \node [rotate=90] {\textcolor{black}{$\Leftarrow$}};    
    \end{tikzpicture}
    };
    \node at (1,-1.3) {\scriptsize $\CB^\pm$};

    \node at (2.3,-4) 
    {\begin{tikzpicture}
      \node [rotate=135] {\textcolor{black}{$\Leftarrow$}};    
    \end{tikzpicture}
    };
    \node at (1.7,-4.2) {\scriptsize $\CA$};

    \node at (8.7,-4) 
    {\begin{tikzpicture}
      \node [rotate=45] {\textcolor{black}{$\Rightarrow$}};    
    \end{tikzpicture}
    };
    \node at (9.5,-4) {\scriptsize $\CB^\mp$};

    \node at (9,-1.3) 
    {\begin{tikzpicture}
      \node [rotate=90] {\textcolor{black}{$\Rightarrow$}};    
    \end{tikzpicture}
    };
    \node at (9.5,-1.3) {\scriptsize $\CA$};

    \node at (5.3,0) 
    {\begin{tikzpicture}
      \node [rotate=0] {\Large \textcolor{black}{$=\!\!=$}};    
    \end{tikzpicture}
    };
    \node at (5.3,-0.5) {\scriptsize $\t=\pm M(1\texttt{-}\D)$};
    
\end{tikzpicture}
\end{equation}
where the adjoint of R-charge $2\d$ in (a) is deconfined via BBP duality, denoted by $\CB^\pm$ according to the magnetic charge of the monopole in the superpotential, into a $U(N-1)$ node with bi-fundamentals of R-charge $\d$, as shown in (b). We then apply Aharony duality $\CA$ to the $U(N)$ node to obtain (c), where the monopole R-charge satisfies
\begin{align}
    (N-1)\d + M \D - M + 2 \mp 
    \big( 
    -\t \pm \d (N-1)
    \big)
    =
    2
    \,,
\end{align}
which allows us to apply $\CB^{\mp}$ successively to obtain (d). We then confine the $U(N-1)$ tail node into an adjoint via $\CA$, yielding (e) with two singlets. Note that the rank of the original $U(N)$ node in (a) decreases by one in (e), while its FI-parameter remains unchanged. We can thus repeat this procedure $N$ times until the gauge rank vanishes, leaving $2N$ singlets:
\begin{equation}
    \begin{tikzpicture}[thick,scale=1.1, every node/.style={scale=0.9}]%[shorten >=1pt]
    \tikzset{vertex/.style={circle,fill=white!25,minimum size=12pt,inner sep=2pt}}
      \tikzset{every loop/.style={}}%removes arrow head from all loops.
        % Theory A
        \node[vertex] (AN) at (0,0) [shape=circle,draw=black,minimum size=2em] {$N$};
        \node[vertex] (AF) at (1.5,0) [shape=circle,draw=black,minimum height=2em, minimum width=2em] {$M$};
        \node[vertex] (DM) at (4.2,0) [shape=circle,draw=black,minimum height=2em, minimum width=2em] {$M$};

        \node at (2.2,0) {$\cdots$};
        \node at (4.9,0) {$\cdots$};
    
        \draw[-to, min distance=1cm]  (AN) edge [out=120, in=60] node {} (AN);
        \draw[->-=.5] ([yshift= 2pt] AN.east) to ([yshift= 2pt] AF.west);
        \draw[->-=.5] ([yshift= -2pt] AF.west) to ([yshift= -2pt] AN.east);
    
        \node at (0,1.2) {\scriptsize $2\d$};
        \node at (0.75,0.4) {\scriptsize $r$};
        \node at (0,-0.6) {\scriptsize $\pm M(1\texttt{-}r)$};
        \node at (1.5,-0.6) {\scriptsize $\k$};
        \node at (4.2,-0.6) {\scriptsize $\k \pm N(1-r)$};

        \node at (3,0) { $=\!\!\!=$};

        \node at (4.5,0.8) {$+ \big\{\; 2\d j  \;\big\}_{j=1-N}^{N} $};
        \node at (6.5,-0.5) {$\blacksquare$};
    \end{tikzpicture}
\end{equation}

\section{{\tt Quiver Lab} : 3d \texorpdfstring{$\CN=2$}{N=2} dualities as quiver diagrams} \label{app: quiver lab}
\begin{figure}[thbp]
\centering
\includegraphics[width=0.95\textwidth]{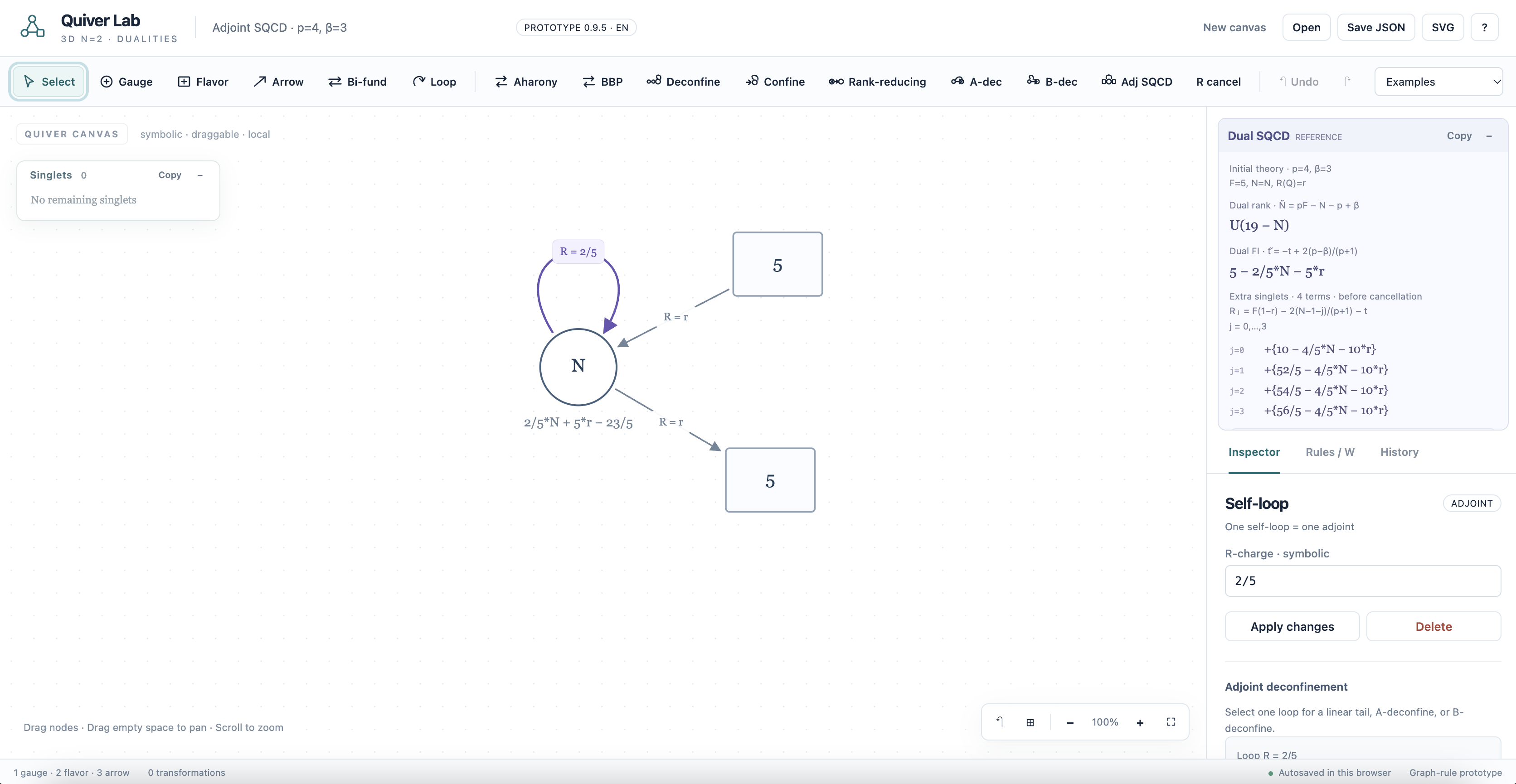}
\caption{\label{fig: quiver lab} Interface of {\tt Quiver Lab}. }
\end{figure}

We provide {\tt Quiver Lab}, a standalone, installation-free HTML application for representing 3d $\CN=2$ ellipsoid partition functions as quiver diagrams and performing duality transformations following the conventions of this paper. The application can be accessed directly through a web browser at \url{https://sungjoonkim-create.github.io/quiver-lab/} and is publicly available at \url{https://github.com/sungjoonkim-create/quiver-lab}. See Figure~\ref{fig: quiver lab} for its interface.

The current version supports only 3d $\CN=2$ gauge theories with unitary gauge groups and implements Aharony and BBP dualities. Its main features are summarized below:
\begin{itemize}
    \item {\tt Gauge/Flavor :}
These create circular/rectangular nodes representing gauge/flavor symmetries, respectively. An FI-parameter must be specified for each gauge node.

\item {\tt Arrow/Bi-fund/Loop :}
These create a directed arrow/a pair of oppositely directed arrows/a self-loop, respectively. An R-charge must be specified for each.

\item {\tt Aharony/BBP :}
These perform $\CA$/$\CB$ duality transformations on the selected node as quiver mutations. The resulting singlet contributions are collected in the box at the upper left of Figure~\ref{fig: quiver lab}. Singlet pairs whose R-charges sum to $2$, as well as individual singlets of R-charge $1$, are automatically removed.

\item {\tt Deconfine/Confine :}
For an adjoint (loop) of R-charge $2/(p+1)$ with $p\in\mathbb{Z}_+$, {\tt Deconfine} replaces it with a length-$p$ $\CC_p[U(N)]$ tail. {\tt Confine} reverses this procedure when the tail has the required structure.

\item {\tt Rank-reducing :}
This removes a gauge node when its FI-parameter satisfies the condition for the rank-reducing duality chain.

\item {\tt A-dec/B-dec :}
These deconfine an adjoint (loop) into an additional gauge node and bi-fundamentals via Aharony/BBP duality, respectively.

\item {\tt Adj SQCD :}
This generates adjoint SQCD with a monopole superpotential term $\hat V_\b^+$. The inputs $(p,F,N,\b)$ follow the conventions of \eqref{eq: the dual}.

\item {\tt R cancel :}
This removes massive contributions from the quiver, including adjoints of R-charge $1$, bi-fundamental pairs of R-charge $1$, and singlet pairs whose R-charges sum to $2$.
    
\end{itemize}

\bibliographystyle{ytphys}
\bibliography{ref}
\end{document}